\documentclass[final,3p,times]{elsarticle}

\usepackage{natbib}
\usepackage{array}
\usepackage{booktabs}
\usepackage{tabu}
\usepackage{dcolumn}
\usepackage{amsmath}
\usepackage{amsfonts}
\usepackage{amssymb}
\usepackage{graphicx}
\usepackage{subfigure}
\usepackage{epsfig}
\usepackage{xcolor}
\usepackage{float}
\usepackage[breaklinks=true,colorlinks=true,linkcolor=blue,urlcolor=blue,citecolor=red]{hyperref}

\usepackage{dcolumn}
\usepackage{bm}

\def\be{\begin{equation}}
\def\ee{\end{equation}}
\def\bea{\begin{eqnarray}}
\def\eea{\end{eqnarray}}

\journal{Physics of The Dark Universe}

\begin{document}

\begin{frontmatter}



\title{Observational constraints on the interacting dark energy - dark matter (IDM) cosmological models }


\author[BBU,DTP]{T. Harko}\ead{tiberiu.harko@aira.astro.ro}


\author[IPM]{K. Asadi}\ead{k.asadi@ipm.ir}

\author[IPM]{H. Moshafi}\ead{moshafi@ipm.ir}

\author[IPM,NWU,CQRC]{H. Sheikhahmadi\corref{cor1}}\ead{h.sh.ahmadi@gmail.com;h.sheikhahmadi@ipm.ir}

\cortext[cor1]{Corresponding author}
\address[BBU]{Department of Physics, Babes-Bolyai University, Kogalniceanu Street,
400084 Cluj-Napoca, Romania,}

\address[DTP]{Department of Theoretical Physics, National Institute of Physics
and Nuclear Engineering (IFIN-HH), Bucharest, 077125 Romania,}

\address[IPM]{School of Astronomy, Institute for Research in Fundamental Sciences (IPM),  P. O. Box 19395-5531, Tehran, Iran}
\address[NWU]{Center for Space Research, North-West University, Potchefstroom, South Africa}
\address[CQRC]{Canadian Quantum Research Center 204-3002 32 Avenue Vernon, British Columbia V1T 2L7, Canada}

\begin{abstract}
 Particle creation (or annihilation) mechanisms, described by either quantum field theoretical models, or by the thermodynamics of irreversible processes, play an important role in the evolution of the early Universe, like, for example, in the warm inflationary scenario. Following a similar approach, based on the thermodynamics of open systems, in the present work we investigate the consequences of the interaction, decay, and particle generation in a many-component Universe, containing dark energy, dark matter,  radiation, and ordinary matter, respectively. In this model, due to the interaction between these components, and of the corresponding thermodynamical properties, the conservation equations of different cosmological constituents are not satisfied individually. We introduce a thermodynamic description of the Universe, in which two novel physical aspects, the particle number balance equations, and the creation pressures, are considered, thus making the cosmological evolution equations thermodynamically consistent. To constrain the free parameters of the model several  observational data sets are employed, including the Planck data sets, Riess 2020, BAO, as well as Pantheon data. By using a scaling ansatz for the dark matter to dark energy ratio, and by imposing constraints from Planck+Riess 2020 data, this model predicts an acceptable value for the Hubble parameter, and thus it may provide a solution to the so-called Hubble tension problem, much debated recently.
\end{abstract}


\end{frontmatter}
\tableofcontents


\
\section{Introduction}\label{Intro00}

The observational data obtained from the investigations of the Type Ia supernovae, presented initially in \citep{Ri98,Ri98-1,Ri98-2,Ri98-3, Ri98-4,Hi,Ri98-5}, and strongly suggesting that the Universe is in an accelerated, de Sitter type,  expansionary state, have led to intensive observational and theoretical efforts for the understanding of the present day cosmological dynamics (for a review of the evidences for cosmic acceleration see \cite{Ri98-6}). Further investigations of  the Cosmic Microwave
Background performed by the Planck satellite \citep{C4}, and the study of the Baryon Acoustic Oscillations \citep{Da1,Da2,Da3} have provided a strong support for the interpretation of the observational data as indicating an accelerating expansion of the Universe. But this remarkable discovery also needs a fundamental change in our current understanding of the gravitational force. The simplest explanation for the accelerated expansion may be obtained by resorting to the cosmological constant $\Lambda$, introduced by Einstein in 1917 \citep{Ein} in order to build a static model of the Universe. However, even if the cosmological constant gives an excellent fit to the observational data, its physical or geometrical nature is still unknown (for extensive reviews of the cosmological constant problem see \citep{W1,W2,W3}.

Hence, in order to solve some of the theoretical problems related to the cosmological constant the existence of a new fundamental component of the Universe, called dark energy, was assumed (for reviews of the dark energy models see \citep{PeRa03, Pa03,DE2,DE3,DE4,DE1}.  The simplest dark energy model consists of a single scalar field $\phi$, in the presence of a self-interaction potential $V(\phi)$. The gravitational action takes the form
\be
S=\int{\left[\frac{M_{p}^2}{2}R-\left(\partial \phi\right)^2-V(\phi)\right]\sqrt{-g}d^4x},
\ee
where $M_p$ denotes the Planck mass, and $R$ is the Ricci scalar. The dark energy models constructed in this way are called quintessence models \citep{quint1,quint2,quint2b,quint3,quint4}.

Another fundamental question in modern cosmology and astrophysics is the dark matter problem
 (see \citep{DMR1,DMR2,DMR3} for detailed reviews of the recent results
on the properties of dark matter, and for its search). The existence of dark matter at a galactic and extragalactic scale is needed for the explanation of two fundamental observations, the dynamics of the galactic rotation curves, and the mass discrepancy in clusters of galaxies, respectively. The galactic rotation curves \citep{Sal,Bin, Per, Bor} indicates that Newtonian gravity as well as standard general relativity may not be able to independently describe galactic dynamics. To explain the galactic rotation curves and the virial mass discrepancy in clusters of galaxies one needs to assume the existence of some dark (invisible) matter, interacting only gravitationally, and which is distributed in a spherically symmetric halo around the galaxies. Dark matter is assumed to form a cold, pressureless cosmic component.

There are many candidates for dark matter that have been proposed, including WIMPs (Weakly Interacting Massive Particles), axions, neutrinos, gravitinos, neutralinos etc. (for reviews of the dark matter candidates see \citep{Ov,Ov1,Ov2,Ov3}. The interaction cross sections of the dark matter particles with normal baryonic matter are assumed to be very small, however, they are expected to be non-zero,  thus making their direct experimental detection possible.
Hence, according to the present view of the cosmological evolution, the dynamics and the expansion of the Universe is dominated by dark energy, and cold dark matter, with baryonic matter playing a minor role in the late time cosmology.

The simplest model that can explain the late time accelerating de Sitter phase is based on the reintroduction in the gravitational field equations of the cosmological constant $\Lambda$.  The $\Lambda$ extension of the Einstein gravitational field equations is the theoretical basis of the present day standard cosmological paradigm, called the $\Lambda$CDM ($\Lambda$Cold Dark Matter) model, in which cold dark matter also plays a fundamental role. Despite its simplicity, the $\Lambda$CDM model provides a very good fit to the cosmological observations \citep{C4, C1,C2,C3}.

On the other hand, the $\Lambda $CDM model also faces some serious (and yet unsolved) observational problems.  The most interesting and important of these problems is the so-called "Hubble tension", which has its origin in the severe differences between the numerical values of the Hubble constant, $H_0$, as determined from the measurements of the Cosmic Microwave Background by the Planck satellite \citep{C4}, and the estimations obtained directly from the astrophysical and cosmological observations in the local Universe \citep{M1,M2,M3,Banerjee:2020xcn}. Thus, for example, the SH0ES (Supernovae and $H_0$ for the Equation of State of Dark Energy) determinations of $H_0$ give the value $H_0 = 74.03 \pm 1.42$ km/s/Mpc \citep{M1}. But from the analysis of the CMB, originating in the early Universe, as performed by the Planck satellite one finds  $H_0 = 67.4 \pm 0.5$ km/s/Mpc \citep{C3}, a numerical value that differs from the SH0ES result by $\sim  5\sigma$ \citep{H0-a}. {For a detailed review of the Hubble tension problem, including a discussion of recent Hubble constant estimates and a summary of the proposed theoretical solutions, see} \citep{Val}. Presently, no specific model can be considered as being  better than all the others. However, proposals and models involving modified gravity, neutrino interactions, early or dynamical dark energy, interacting cosmologies, or primordial magnetic fields, may provide some good options and approaches to the problem, until a better alternative is found.

Dark energy has an antigravitational effect, determining the Universe to accelerate at late times. On the other hand, dark matter experiences only  gravitational attraction. In the $\Lambda$CDM approach to cosmology no interaction between these two components is allowed. The gravitational properties of the dark matter and of the dark energy are contrary (gravitational attraction versus gravitational repulsion). Moreover, dark energy is homogeneously distributed through the entire Universe, while dark matter gravitates around baryonic matter. Hence, one could suppose that any interaction between these two fundamental components of the Universe would be negligible, or extremely weak.

However, the possibility of a dark energy-dark matter interaction cannot be excluded {\it a priori}.  Some early proposals in this direction were formulated initially in  \citep{int,int1,int2,int3,int4}, and later on they developed into an active field of research in cosmology and astrophysics \cite{lit} - \cite{Yang:2018euj}.
 For a review of the dark matter and dark energy interactions, including the  present day theoretical challenges, the cosmological implications, and the observational signatures see \citep{revDEDM}.

In the standard approach to the dark energy - dark matter interaction problem, dark energy is modelled as a scalar field with energy density $\rho_{\phi }$ and pressure $p_{\phi }$, respectively, while dark matter is considered as a fluid with density $\rho _{DM}$ and pressure $p_{DM}$, respectively, obeying an equation of state of the form $w_{DM}=p_{DM}/\rho_{DM}\neq 0$. By assuming for the cosmological background an isotropic and homogeneous  Friedman-Lemaitre-Robertson-Walker (FLRW)
geometry with scale factor $a(t)$, and assuming that the creation/annihilation of the scalar field (dark energy) and of the
dark matter fluid occurs at a rate $Q$, the basic equations describing the variations of the densities of the dark energy $\rho _{\phi }$ and of the dark matter $\rho _{DM}$ are given by \citep{int,lit}
\bea\label{1}
\dot{\rho _{\phi }}+3H\left(1+w_{\phi}\right)\rho _{\phi}&=&-Q,
    \\
\label{2}
\dot{\rho} _{DM}+3H\left(1+w_{DM}\right)\rho _{DM}&=&+Q,
\eea
respectively, where $w_{\phi}=p_{\phi}/\rho _{\phi}$, and $H \equiv \dot{a}/a$ is the Hubble function. In the above and the following equations an overdot indicates the derivatives with respect to the cosmological time $t$. Hence, in the above model the dark energy  and the dark matter transform into each other, with the creation/decay processes described by a single creation/decay rate $\pm Q$. Therefore, the function $Q$ describes the dynamical interaction at a cosmological level between the two fundamental dark components of the Universe.  If $Q > 0$, it turns out that dark energy is converted into dark matter. In the opposite case, with $Q < 0$, dark matter is converted into dark
energy.

Since up to now no rigorous theoretical approach that could predict the functional form of the function $Q$, describing the coupling between dark matter and dark energy, does exist, the interacting DE-DM  models are essentially phenomenological. However, some couplings may be considered more natural,  or more physical, than others. Therefore, a large number of functional forms for $Q$ have been assumed a priori, and investigated in detail in the literature. Some interesting forms of $Q$ are $Q=\rho _{crit}^0(1+z)^3H(z)I_Q(z)$, where $z$ is the redshift, and $I_Q(z)$ is an interaction function that depends on the redshift \citep{Cueva}, or $Q\propto \rho _{DM}\dot{\phi }$, and $Q\propto H\rho _{DM}$, respectively \citep{Singl}.

An interacting Dark Matter-Dark Energy  model was investigated in \citep{lit38},  by simulating the luminosity distance for realistic GW+GRB events, which can be detected by the third-generation GW detector of the Einstein Telescope. Using these simulated events,  a Monte Carlo Markov Chain analysis was performed to constrain the DM-DE coupling constant and other model parameters at 1$\sigma$ and 2$\sigma$ confidence levels. A model of dark matter-dark energy interaction, with coupling strength proportional to the multiplication of dark sector densities with different power indices, was investigated, and constrained, in \citep{lit39}. Overall, the DM-DE interaction model is consistent with the current observational data, and provides a better fit to the redshift distortion data. The effects of DE-DM interaction on the ellipticity of cosmic voids was studied in \citep{lit40}, by applying the linear growth of density perturbation in the presence of interaction. It was shown that the ellipticity of cosmic voids increases when the DM and DE interaction is considered. A specific form of the coupling between DE and DM, proportional to the dark energy density, was investigated in \citep{lit41}. Once all relevant cosmological probes are included simultaneously, the value of the Hubble parameter in this model is $H_0=69.82_{-0.76}^{+0.63}$ km /(s Mpc ), which reduces the Hubble tension to 2.5 $\sigma$.

A cosmological scenario where the dark sector is described by two perfect fluids that interact through a velocity-dependent coupling was considered in \citep{lit42}. The interaction of both dark components leads to a suppression of the dark matter clustering at late times. This suppression of clustering, together with the additional dark radiation at early times, can simultaneously alleviate the $S _8$ and $H_0$ tensions. The dynamics of the interacting dark energy and dark matter in viable models of $f(R)$ gravity was investigated in \citep{lit43} by using a standard framework of dynamical system analysis. The fixed points are slightly modified from those obtained in the standard non-interacting analysis of $f(R)$ gravity. The dynamical profiles of the Universe in the viable $f(R)$ dark energy models are modified by the interaction term, together with their model parameters. A scenario where the dark energy is a dynamical fluid whose energy density can be transferred to the dark matter via a coupling function proportional to the energy density of the dark energy was considered in \citep{lit44}. Against data from Planck, BAO and Pantheon, the model can significantly reduce the importance of the $S_8$ tension,  without exacerbating nor introducing any other tension (such as the $H_0$ tension), and without worsening the fit to the considered data sets, with respect to the $\Lambda $CDM model.

In postulating Eqs.~(\ref{1}) and (\ref{2}) to describe the dark energy-dark matter interaction, one adopts a hypothetical picture of a very rapid decay of dark energy into dark matter, in which the dark matter particles reach an equilibrium state immediately after dark matter particles are produced. But from a general physical point of view it is reasonable to assume that during the dark energy/dark matter creation/annihilation period complicated nonequilibrium processes may also occur, and that the newly produced particles have a highly nonequilibrium distribution. The relaxation to an equilibrium state may therefore occur during a long period of cosmological evolution.

Thermodynamical systems in which particle creation occurs belong to the important class of open thermodynamical systems, in which the usual adiabatic laws of the conservation of particle number and energy are adapted to include irreversible particle creation \citep{Prig}. The resulting thermodynamic formalism of open systems in the presence of particle production has found a large number of applications in cosmology, where it was first applied to obtain a specific cosmological model in \citep{Prig}. By explicitly including particle creation related terms in the total matter energy-momentum tensor in the Einstein gravitational field equations one obtains a three phases cosmological model that begins with an instability of the vacuum. Then the Universe is driven from the initial fluctuation of the vacuum to a de Sitter type expansionary phase, during which  particle creation occurs. During the annihilation of its components the Universe exists in the de Sitter phase (second stage),
and enters, after a phase transition, into the isotropic and homogeneous, ordinary matter dominated FLRW Universe.

The initial thermodynamic approach introduced in \citep{Prig} was generalised in \citep{Cal}, where a covariant formulation of irreversible thermodynamic was developed, in which specific entropy variations, usually appearing in non-equilibrium processes, where also included. Cosmological models based on the irreversible thermodynamics of open systems in the presence of particle creation have been considered in \cite{cosm} - \cite{cosm21}.
In \citep{cosm10a} it was pointed out that in modified gravity theories with geometry - matter coupling, in which the action is an arbitrary function of the Ricci scalar and of the matter Lagrangian $(f\left(R,L_m\right) gravity)$, or of the Ricci scalar and of the trace of the matter energy-momentum tensor ($f(R,T)$ gravity), in which the divergence of the matter energy - momentum tensor is nonzero, by using the formalism of open thermodynamic systems, the generalized conservation equations in these gravitational theories can be interpreted, from a thermodynamic point of view, as describing irreversible matter creation processes. Matter production thus corresponds to an irreversible energy flow from the gravitational field to the newly created particles. Matter creation processes during the reheating period at the end of inflation in the early Universe can also be described by using the irreversible thermodynamic of open systems \citep{cosm12a}. The particle content of the very early Universe is assumed to consist of the inflationary scalar field, which, through its decay, generates relativistic ordinary matter, in the form of radiation, and pressureless dark matter, respectively. At the early stages of reheating the inflationary scalar field transfers its energy to the newly created particles, with the field energy decreasing to near zero.  The thermodynamics of open systems was applied, together with the gravitational field equations, to a two-component cosmological fluid consisting of a scalar field and dark matter, in \citep{lit19}.  Hence a generalization of the elementary dark energy-dark matter interaction theory was obtained, in which the decay (creation) pressures are explicitly considered as components of the energy-momentum tensor of the cosmological fluid. In the framework of warm inflationary models the interaction between scalar fields and radiation was modeled by using the irreversible thermodynamics of open systems with particle creation/annihilation in \citep{Harko:2020cev}.

It is the main goal of the present paper to consider the applications of the irreversible thermodynamics of open systems, as introduced in \citep{Prig}, and further developed in \citep{Cal}, respectively, to a cosmological fluid mixture consisting of three distinct components: dark energy, described by a scalar field, dark matter, and radiation, modeled as an ordinary matter fluid, respectively. We assume that in this system particle decay and production does occur through three channels, involving the decay of the scalar field, and the creation of dark matter, and ordinary matter, respectively. Such a physical framework may describe both the early and the late stages of cosmological evolution.  The thermodynamics of irreversible processes and open systems as applied to this three component cosmological model, with interacting scalar field dark energy, and  matter, with dark and ordinary types, leads to a self-consistent representation of the dark energy and particle creation/annihilation processes, which in turn decide the dynamics, and the present and future evolution of the Universe, respectively.

The present paper is organised as follows. In Section~\ref{sect2} we present, in some detail, due to its important role, the thermodynamical theory of irreversible matter creation processes. The theory is applied to a three-component cosmological fluid with interacting dark energy, dark matter and ordinary matter, or briefly interacting dark energy-matter (IDM) model.  The resulting gravitational field equations are written down in Section~\ref{ThreeComp01}. Particular models, and exact and numerical solutions to the field equations are considered in Section~\ref{ThreeComp02}. In Section~\ref{MOD}, besides determining the methodology of observational constraints, we obtain observational constraints on the free parameters of the model, and investigate the concordance between the theoretical predictions of the model, and observations. In Section~\ref{Conclustions} we discuss and conclude our results. Throughout the paper we use a system of units so that $8\pi G=c=1$.
\section{Thermodynamical description of cosmological particle creation}\label{sect2}
In the present Section we will briefly review the fundamentals of the thermodynamics of open systems and the basic relation to be used in the sequel.
We begin our analysis of particle creation by considering a large cosmological volume element $V$ that contains $N$ interacting particles. Let the internal energy of the system be $E$. If the system is a closed thermodynamical one,  $N$ is constant. The conservation of the internal energy $E$ is the expressed by the first law of thermodynamics according to the fundamental relation $dE=dQ-pdV$ \citep{Prig} where $dQ$ denotes the heat transferred to the system during time $dt$, $V$ is any comoving volume, and $p$ is the thermodynamic pressure. We introduce now the energy density $\rho$, defined as $\rho = E/V$, the heat per unit particle $dq$, with $dq = dQ/N$, and the particle number density $n$, given by $n = N/V$. Then the first law of thermodynamics takes the form
\be\label{encons1}
d\left(\frac{\rho }{n}\right)=dq-pd\left(\frac{1}{n}\right).
\ee

It is important go note at this moment that Eq.~(\ref{encons1}) is also true for open systems, in which the particle number $N$ is time dependent, so that $N=N(t)$.

\subsection{Covariant formulation of particles decay/creation processes}

The basic macroscopic quantities that fully characterize the thermodynamic states of a relativistic fluid are its energy-momentum tensor $T_{\mu \nu}$, the particle flux vector $N^{\mu }$, and the entropy flux vector $s^{\mu }$, respectively. The energy-momentum tensor $T_{\mu \nu}$ satisfies the covariant conservation law $\nabla _{\nu }T^{\mu \nu}=0$, where $\nabla _{\mu}$ denotes the covariant derivative with respect to the metric. By taking into account matter creation the energy-momentum tensor can be written as
 \be\label{tmunu}
 T^{\mu \nu}=\left(\rho +p+p_c\right)u^{\mu }u^{\nu }-\left(p+p_c\right)g^{\mu \nu},
 \ee
where $u^{\mu}$  is the four-velocity of the fluid, normalized according to the relation $u^{\mu}u_{\mu}=1$,  while the creation
pressure $p_c$ characterizes phenomenologically particle decay/creation, as well as other possible dissipative thermodynamic effects.

The particle flux vector is defined according to $N^{\mu} =nu^{\mu }$, and it satisfies the balance equation $\nabla _{\mu}N^{\mu }=\Psi$,
where the function $\Psi $ is a matter source if $\Psi >0$, and a matter sink for $\Psi <0$. In the standard approaches to cosmology $\Psi $ is usually assumed to be zero, and therefore no particle decay/production processes are considered. The entropy flux $s^{\mu }$ is defined as $s^{\mu }=n\sigma u^{\mu }$  \citep{Cal}, where by $\sigma $ we have denoted the specific entropy per particle. The second law of thermodynamics imposes the condition $\nabla _{\mu }s^{\mu }\geq 0$. Another important thermodynamic relation, the Gibbs equation, is given for an open thermodynamic system with temperature $T$ in the presence of particle creation by \citep{Cal}
\be
nTd\sigma =d\rho -\frac{\rho +p}{n}dn.
\ee

The entropy balance equation can be immediately obtained by using the above equations as \citep{Cal}
\be\label{eq1}
\nabla _{\mu }s^{\mu }=-\frac{p_c \Theta }{T}-\frac{\mu \Psi}{T},
\ee
where $\Theta =\nabla _{\mu }u^{\mu}$ denotes the expansion of the fluid, while the chemical potential $\mu $ is given by Euler's relation $\mu =\left(\rho +p\right)/n-T\sigma$.

In the following we assume that in the space-time the newly created particles are in thermal equilibrium with the already
existing ones. Then it follows that the entropy production is due only to the particle creation, which is the dominant process describing the entropy evolution. Moreover, for the creation pressure  $p_c $ we adopt the following phenomenological ansatz \citep{Prig, Cal}
\be
p_c =-\alpha \frac{\Psi}{\Theta},
\ee
where $\alpha >0$. With this choice for the entropy balance we obtain the equations
\be\label{eb}
\nabla _{\mu }s^{\mu }=\frac{\Psi}{T}\left(\alpha -\mu \right)=\Psi \sigma +\left(\alpha -\frac{\rho +p}{n}\right)\frac{\Psi}{T}=\Psi \sigma +n\dot{\sigma },
\ee
where $\dot{\sigma}=u^{\mu }\nabla _{\mu }\sigma $. Together with Eq.~(\ref{eq1}), Eq.~(\ref{eb})  gives for the specific entropy production $\sigma$ the relation \citep{Cal}
\be\label{eq2}
\dot{\sigma }=\frac{\Psi}{nT}\left(\alpha -\frac{\rho +p}{n}\right).
\ee

If we restrict the present thermodynamic formalism by requiring that the specific entropy per particle $\sigma $ is constant, $\sigma ={\rm constant}$, then Eq.~(\ref{eq2}) gives for $\alpha $ the expression $\alpha =\left(\rho +p\right)/n$. Thus, for the creation pressure we obtain the expression \citep{Cal}
\be\label{pc}
p_c=-\frac{\rho +p}{n\Theta }\Psi .
\ee

By imposing the condition of the constancy of the specific entropy, the Gibbs equation takes the form
\be\label{ad}
\dot{\rho }=\left(\rho +p\right)\frac{\dot{n}}{n}.
\ee

\subsection{Matter creation in homogeneous and isotropic cosmological models}

In the following we will assume that the geometry of the space-time is homogeneous and isotropic. Moreover, to describe the physical processes in the considered geometry we adopt a comoving frame in which the components of the four-velocity are given by $u^{\mu }=\left(1,0,0,0\right)$. Furthermore, we assume that the geometric as well as the thermodynamic quantities are a function of the time $t$ only. Then, the covariant derivative of any function $f(t)$, constructed as $\dot{f}=u^{\mu}\nabla _{\mu}f$  coincides with the ordinary time derivative, $\dot{f}=u^{\mu }\nabla _{\mu }f=df(t)/dt$. The expansion of the fluid is obtained as $\nabla _{\mu }u^{\mu }=\dot{V}/V$, where $V$ is the comoving volume element.

Eq.~(\ref{ad}) can be represented in a number of equivalent forms according to
\be\label{rhodot}
\dot{\rho }=\left(\frac{h}{n}\right)\dot{n},
\ee
where  $h = \rho + p$ denotes the enthalpy (per unit volume) of the fluid, or, equivalently,
\be
p=\rho \left(\frac{\dot{\rho }}{\rho}-\rho \frac{\dot{n}}{n}\right).
\ee

The Einstein gravitational field equations
\be
R_{\mu \nu }-\frac{1}{2}g_{\mu \nu }R=T_{\mu \nu },
\ee
contain the macroscopic energy-momentum tensor $T_{\mu \nu }$. In
the standard cosmological case, for $T_{\mu \nu }$ one adopts to form corresponding to a perfect fluid. In the presence of matter creation the energy-momentum tensor takes a similar form, and it is characterised by a phenomenological energy density $\rho $, and pressure $\bar{p}=p+p_c$, which includes the effects of particle decay/creation. In the comoving frame the components of $T_{\mu \nu}$ are given by
\be
T_0^0=\rho, T_1^1=T_2^2=T_3^3=-\bar{p}.
\ee
The Einstein field equations imply, via the geometric Bianchi identities, the condition $\nabla _{\nu }T_{\mu }^{\nu }=0$, leading to the relation
\be\label{cons2}
d(\rho V ) = -\bar{p}dV.
\ee

In the presence of adiabatic irreversible matter decay/creation processes the analysis of physical processes
must be performed by using the thermodynamics of open systems. Hence one must consider an effective pressure that includes the
supplementary decay/creation pressure $p_c$. Hence Eq.~(\ref{ad}) can be written in a form similar to Eq.~(\ref{cons2}), namely \citep{Prig}
\be\label{cons3}
d(\rho V ) = -\left(p+p_c\right)dV=-\bar{p}dV,
\ee
where $\bar{p} = p + p_c$. Then from Eq.~(\ref{pc}) it follows that the creation pressure $p_c$ can be obtained as
\be\label{pc1}
p_c = -\left(\frac{h}{n}\right)\frac{d(nV)}{dV}=-\left(\frac{h}{n}\right)\frac{V}{\dot{V}}\left(\dot{n}+\frac{\dot{V}}{V}n\right).
\ee

The creation of particles corresponds to a (negative) supplementary pressure $p_c$, which needs to be considered as an independent component of the total cosmological pressure $\bar{p}$ that enters in the Einstein gravitational field equations (the decay of matter corresponds to a positive annihilation pressure).

The entropy variation $dS$ in an open thermodynamic system can be decomposed into two components, the entropy flow $d_0S$, and the entropy creation term $d_iS$, respectively, so that $dS = d_0S + d_iS$, with $d_iS \geq 0$. To calculate $dS$ we consider the total differential of the entropy, given by $Td(sV ) = d(\rho V ) + pdV -\mu d(nV )$,
where $s= S/V\geq 0$, $\mu n = h - Ts$, and $\mu \geq 0$. Since in a homogeneous system $d_0S = 0$, it follows that only particle creation contributes to the entropy production. Hence for the rate of the entropy production we immediately find \citep{Prig}
\be
T\frac{dS}{dt}= T\frac{d_iS}{dt}= T\frac{s}{n}\frac{d(nV)}{dt}.
\ee

To close the problem we require one more relation between the particle number $n$ and the comoving volume $V$, describing the evolution of $n$ as a result of particle creation/annihilation processes. This relation can be obtained from the equation $\nabla _{|mu}N^{\mu}=\Psi$, which in the case of a homogeneous and isotropic cosmological model gives
\be\label{Psi}
\frac{1}{V}\frac{d(nV )}{dt}=\Psi (t),
\ee
where $\Psi(t)$ is the matter creation (or annihilation) rate ($\Psi (t) > 0$ indicates particle creation, while $\Psi (t) < 0$ corresponds to particle annihilation) \citep{Prig,Cal}. The creation pressure (\ref{pc}) is a function of the particle creation (annihilation) rate, and hence it couples Eqs.~(\ref{pc}) and (\ref{Psi}) to each other, and, notwithstanding indirectly, both of them with the total energy conservation equation (\ref{cons3}), which is a consequence of the Einstein gravitational field equations themselves. 

\section{Cosmological models with three interacting components in the late\\
 Universe: a general approach}\label{ThreeComp01}

In the present Section we consider a cosmological model of a Universe consisting of three distinct, but interacting components: dark energy (DE), dark matter (DM), and ordinary matter (OM), respectively. Each of these components are characterized by their energy densities $\rho_{DE}$, $\rho_{DM}$, $\rho_{OM}$, and by their pressures $p_{DE}$,$p_{DM}$, and $p_{OM}$, respectively. Dark energy decays into dark and ordinary matter, thus leading to the particle number variation of each component. Furthermore, we assume that the decay rate of the dark energy is proportional to its energy density $\rho_{DE}$, and that the creation rates of the dark and normal matter are also proportional to $\rho _{DE}$. From a physical point of view such a system can be described by using the thermodynamics of open systems. As for the geometry of the Universe, we assume that it is described by the homogeneous and isotropic Friedmann-Lemaitre-Robertson-Walker metric,
\be
ds^2=dt^2-a^2(t)\delta_{ij}dx^idx^j,
\ee
where $a(t)$ is the scale factor. We also introduce the Hubble parameter $H(t)$ defined as $H =\dot{a}(t)/a(t)$.

\subsection{Energy balance, creation pressure, and creation rates}

The balance equations for a three component Universe with decaying dark energy and creation of dark and normal matter are given by, \citep{Harko:2020cev}
\begin{eqnarray}\label{conservationDE}
	\dot{n}_{DE} + 3Hn_{DE} = -\Gamma \rho_{DE}\,,
\end{eqnarray}
\begin{eqnarray}\label{conservationDM}
	\dot{n}_{DM} + 3Hn_{DM} = \Gamma_{1} \rho_{DE}\,,
\end{eqnarray}
and
\begin{eqnarray}\label{conservationRAD}
	\dot{n}_{OM} + 3Hn_{OM} = \Gamma_{2} \rho_{DE}\,,
\end{eqnarray}
where the coefficients  $\Gamma$, $\Gamma _1$ and $\Gamma _2$ will be assumed in the following to be constants. By adding the above equations we obtain the total particle number balance as
\be
\dot{n}+3Hn=\left(\Gamma _1+\Gamma _2-\Gamma \right)\rho_{DE},
\ee
where $n=n_{DE}+n_{DM}+n_{OM}$ is the total particle number. If $\Gamma _1+\Gamma _2=\Gamma$, the total particle number is conserved, $\dot{n}+3Hn=0$, but particle creation/decay processes take place in the system.

The creation pressure for this set up can be obtained as
\begin{eqnarray}
	p^{(c)} = -\frac{h}{n} \frac{1}{3H} (\dot{n}+3Hn).
\end{eqnarray}
Hence,  the creation pressures for the components $\left({\rm DE, DM, OM}\right)$ are given by
\begin{eqnarray}\label{pressureDE}
	p^{(c)}_{DE} =\frac{\Gamma(\rho_{DE}+p_{DE})\rho_{DE}}{3H n_{DE}}\,,
\end{eqnarray}
\begin{eqnarray}\label{pressureDM}
		p^{(c)}_{DM} =- \frac{\Gamma_{1}(\rho_{DM}+p_{DM})\rho_{DE}}{3H n_{DM}}\,,
\end{eqnarray}
and
\begin{eqnarray}\label{pressureRAD}
		p^{(c)}_{OM} = -\frac{\Gamma_{2}(\rho_{OM}+p_{OM})\rho_{DE}}{3H n_{OM}}\,.
\end{eqnarray}

By taking into account that the pressure is an additive quantity, by considering Eqs. \eqref{conservationDE} and \eqref{pressureRAD} the total creation pressure equation reads
\begin{eqnarray}\label{Creation-pressure-total}
	p^{(c)}_{tot} =\frac{\rho_{DE}}{3H}\Big[\frac{\Gamma(\rho_{DE}+p_{DE})}{ n_{DE}}- \frac{\Gamma_{1}(\rho_{DM}+p_{DM})}{n_{DM}}-\frac{\Gamma_{2}(\rho_{OM}+p_{OM})}{n_{OM}}\Big]\,.
\end{eqnarray}

For a FLRW metric, the Friedmann equations for the cosmological mixture are given by
\begin{eqnarray}\label{RW01}
	3H^2 = \frac{1}{M_{Pl}^2}\big(\rho_{DE}+\rho_{DM}+\rho_{OM}\big)\,,
\end{eqnarray}
\begin{eqnarray}\label{RW02}
	-2\dot{H}-3H^2 = \frac{1}{M_{Pl}^2}\big[p_{DE}+p_{DM}+p_{OM}+p^{(c)}_{tot}\big]\,.
\end{eqnarray}

In a comoving frame, after taking into account the condition of the constancy of the specific entropy, the Gibbs equation takes the form of Eq.~(\ref{rhodot}).  Hence, for the three component system $\left({\rm DE,DM,OM}\right)$ in the presence of irreversible processes, the energy balance equations are given by  \citep{Harko:2020cev}
\begin{eqnarray}\label{BalanceDE}
\dot{\rho}_{DE} + 3H(1+w_{DE})\rho_{DE} = -\Gamma(1+w_{DE})\frac{\rho_{DE}^2}{n_{DE}}\,,
\end{eqnarray}
\begin{eqnarray}\label{BalanceDM}
	\dot{\rho}_{DM} + 3H(1+w_{DM})\rho_{DM} = \Gamma_{1}\frac{\left(1+w_{DM}\right)\rho_{DM}\rho_{DE}}{n_{DM}}\,,
\end{eqnarray}
and
\begin{eqnarray}\label{BalanceOM}
\dot{\rho}_{OM} + 3H(1+w_{OM})\rho_{OM} =\Gamma_{2}\frac{\left(1+w_{OM}\right)\rho_{OM}\rho_{DE}}{n_{OM}}\,,
\end{eqnarray}
where $w_{DE}$ is the equation of state (EoS) parameter of DE, $w_{DE}=p_{DE}/\rho_{DE}$, $w_{DM}=p_{DM}/\rho_{DM}$, and $w_{OM}=p_{OM}/\rho_{OM}$, respectively.
The above expressions for the energy density of each component can be written as
\begin{eqnarray}\label{Conservation-NEW-DE}
	\dot{\rho}_{DE} + 3H(1+w_{DE})\rho_{DE} = -3 H\Pi_{DE}\,,
\end{eqnarray}
\begin{eqnarray}\label{Conservation-NEW-DM}
	\dot{\rho}_{DM} + 3H(1+w_{DM})\rho_{DM} = -3H\Pi_{DM}\,,
\end{eqnarray}
and
\begin{eqnarray}\label{Conservation-NEW-OM}
	\dot{\rho}_{OM} + 3H(1+w_{OM})\rho_{OM} =-3 H\Pi_{OM}\,,
\end{eqnarray}
where $\Pi_i$, $i={\rm DE,DM,OM}$ are the interaction terms determined as
\be\label{Q-Pis}
\Pi_{DE}=\Gamma(1+w_{DE})\frac{\rho_{DE}^2}{3Hn_{DE}}, \Pi_{DM}=-\Gamma_{1}\frac{\left(1+w_{DM}\right)\rho_{DM}\rho_{DE}}{3Hn_{DM}}, \Pi_{OM}=-\Gamma_{2}\frac{\left(1+w_{OM}\right)\rho_{OM}\rho_{DE}}{3Hn_{OM}}.
\ee

Without loss the generality, in the following we assume
\begin{eqnarray}\label{PIs-multiplying}
	\Pi_{DM} = \alpha \Pi_{OM} \,\,,\,\, \Pi_{DE} = \beta \Pi_{DM} \,\,,\,\, \rho_{OM} = \gamma\rho_{DM},
\end{eqnarray}
where the free parameters $\alpha$, $\beta$ and $\gamma$ will be constrained using different observational data sets .

To derive a specific expression for the interaction terms let us calculate the time evolution of the ratio ${\rho_{DM}}/{\rho_{DE}}$, that reads
\begin{eqnarray}\label{DMtoDE=ratiodot}
	\frac{d}{dt}\left(\frac{\rho_{DM}}{\rho_{DE}}\right) = 3H\Bigg[w_{DE} - \frac{\rho_{tot}}{\rho_{DM}\rho_{DE}}\Pi_{DM}+(1+\gamma+\beta)\frac{\Pi_{DM}}{\rho_{DE}}\Bigg]\frac{\rho_{DM}}{\rho_{DE}}\,.
\end{eqnarray}
To obtain the above expression \eqref{DMtoDE=ratiodot}, we have used Eqs.~(\ref{PIs-multiplying}).

Following \citep{Pavon:2004xk,Zimdahl:2002zb}, we look for solutions with the scaling behavior
\begin{eqnarray}\label{ScalingExpression}
	\frac{\rho_{DM}}{\rho_{DE}} \equiv r \left(\frac{a_{0}}{a}\right)^{\xi} \,	=r(1+z)^{\xi},
\end{eqnarray}
where $r$ denotes the ratio of two components (DM and DE) at the present time , $a = a_0$, and the parameter $\xi$ is another free parameter of the model. In addition, $z$ denotes the redshift defined according to $1 + z\equiv a_0/a$. Then, with the use of \eqref{ScalingExpression} and \eqref{PIs-multiplying} in \eqref{DMtoDE=ratiodot} an expression for $\Pi_{DM}$ can be obtained as
\begin{eqnarray}\label{PI-DM}
	\Pi_{DM} = \Bigg(\frac{\xi+3w_{DE}}{3}\Bigg)\Bigg[\frac{1}{1-r\beta(1+z)^{\xi}}\Bigg]\rho_{DM}\,.
\end{eqnarray}

With the use of  Eqs.~\eqref{PIs-multiplying} and \eqref{ScalingExpression}, the functions $\Pi_{DE}$ and $\Pi_{OM}$ can be obtained immediately. By considering Eqs. \eqref{BalanceDE}, \eqref{BalanceDM} and \eqref{BalanceOM}, the particle number density for each component is given by
\begin{eqnarray}
	n_{DE} = \frac{\Gamma(1+w_{DE}) \left[1-r\beta(1+z)^{\xi}\right]}{Hr\beta\left(\xi+3w_{{DE}}\right)(1+z)^{\xi}}\rho_{DE}\,,
\end{eqnarray}

\begin{eqnarray}
	n_{DM} = -\frac{\Gamma_{1}\left[1-r\beta(1+z)^{\xi}\right]}{Hr\left(\xi+3w_{{DE}}\right)(1+z)^{\xi}}\rho_{DM}\,,
\end{eqnarray}
and
\begin{eqnarray}
	n_{OM} = -\frac{4}{3}\frac{\Gamma_{2}\alpha \left[1-r\beta(1+z)^{\xi}\right]}{Hr\left(\xi+3w_{{DE}}\right)(1+z)^{\xi}}\rho_{OM}\,.
\end{eqnarray}

Now we can  calculate the energy densities of the components of the model.  To do so, we substitute first \eqref{PI-DM} and \eqref{PIs-multiplying} into \eqref{Conservation-NEW-DE}-\eqref{Conservation-NEW-OM}. Then, by integrating the resulting equations we obtain the explicit expressions for the energy densities of DE, DM and OM as follows,
\begin{eqnarray}\label{rhoDE01-G}
	\rho_{DE} = {\rho_{DM0}}\Bigg[\frac{-r\beta+1}{-r\beta(1+z)^{\xi}+1}\Bigg]^{\frac{\xi+3w_{DE}}{\xi}}\frac{(1+z)^{3w_{DE}+3}}{r},
\end{eqnarray}

\begin{eqnarray}\label{rhoDM01-H}
	\rho_{DM} = \rho_{DM0}\Bigg[\frac{-r\beta+1}{-r\beta(1+z)^{\xi}+1}\Bigg]^{\frac{\xi+3w_{DE}}{\xi}}(1+z)^{3w_{DE}+\xi+3},
\end{eqnarray}
and
\begin{eqnarray}\label{rhoOM01-K}
	\rho_{OM} = {\rho_{OM0}}\Bigg[\frac{-r\beta+1}{-r\beta(1+z)^{\xi}+1}\Bigg]^{\frac{\xi+3w_{DE}}{\alpha\gamma\xi}}(1+z)^{\frac{{3w_{DE}+4\alpha\gamma+\xi}}{\alpha\gamma}},
\end{eqnarray}
where the subscript $0$ refers to the present epoch.

\subsection{Dark energy EOS, and the cosmological parameters}

To constrain the free parameters of the model, we need to know the EoS of DE that appears in the above equations. In this regard, we introduce an ansatz for the dark energy EoS parameter as
\begin{eqnarray}\label{Ansatz-EoS}
	w_{DE} = w_{1}(1+z)^{\lambda} + w_{0} - 1\,,
\end{eqnarray}
where $w_{0}$, $w_{1}$, and $\lambda$ are free parameters of the model. By taking $\lambda = 0$, the model reduces to the constant EoS model ($\Lambda$CDM) for the dark energy. By introducing \eqref{Ansatz-EoS} into \eqref{rhoDE01-G}-\eqref{rhoOM01-K},  the energy densities of the components are now obtained as,	

\begin{eqnarray}\label{rhoDE02-GG}
	\rho_{DE} = {\rho_{DM0}}\Bigg[\frac{-r\beta+1}{-r\beta(1+z)^{\xi}+1}\Bigg]^{Y}\frac{(1+z)^{3w_{0}}}{r} A_1(z)\,,
\end{eqnarray}

\begin{eqnarray}\label{rhoDM02-HH}
	\rho_{DM} = \rho_{DM0}\Bigg[\frac{-r\beta+1}{-r\beta(1+z)^{\xi}+1}\Bigg]^{Y}(1+z)^{3w_{0}+\xi} A_1(z)\,,
\end{eqnarray}

\begin{eqnarray}\label{rhoOM02-KK}
	\rho_{OM} = {\rho_{OM0}}\Bigg[\frac{-r\beta+1}{-r\beta(1+z)^{\xi}+1}\Bigg]^{Y/\alpha\gamma}(1+z)^{\chi} A_2(z)\,,
\end{eqnarray}
where $Y$ and $\chi$ are given by the following expressions
\begin{eqnarray}\label{Y01}
		Y = \frac{3w_{0}+\xi-3}{\xi},
\end{eqnarray}
\begin{eqnarray}\label{Xi01}
	\chi = \frac{3w_{0}+4\alpha\gamma+\xi-3}{\alpha\gamma},
\end{eqnarray}
and $A_{1,2}$ are defined as
\begin{eqnarray}\label{A1-01}
A_1(z) = \exp\Bigg[\frac{3 {\omega_1} (1+z)^{\lambda }}{\lambda} \, _2F_1\left(1,\frac{\lambda }{\xi },\frac{\lambda +\xi }{\xi };r (1+z)^{\xi } \beta \right)-\frac{3 {\omega_1}{}}{\lambda} \, _2F_1\left(1,\frac{\lambda }{\xi },\frac{\lambda +\xi }{\xi };r  \beta \right)\Bigg],
\end{eqnarray}
and
\begin{eqnarray}\label{A2-01}
A_2(z) = \exp\Bigg[\frac{3 {\omega_1} (1+z)^{\lambda }}{\alpha \gamma \lambda} \, _2F_1\left(1,\frac{\lambda }{\xi },\frac{\lambda +\xi }{\xi };r (1+z)^{\xi } \beta \right)-\frac{3 {\omega_1}{}}{\alpha \gamma \lambda} \, _2F_1\left(1,\frac{\lambda }{\xi },\frac{\lambda +\xi }{\xi };r  \beta \right)\Bigg],
\end{eqnarray}
respectively, where $_2F_1$ denotes the hypergeometric function $_2F_1(a,b;c;z)=\sum_{k=0}^{\infty}{\left((a)_k(b)_k/(c)_k\right)z^k/k!}$.

Then the total energy density of the Universe is expressed as
\begin{eqnarray}\label{rhoTotal01}
	\rho_{tot}& =&\rho_{m0}(1+z)^{3w_{0}}\Bigg[\frac{-r\beta+1}{-r\beta(1+z)^{\xi}+1}\Bigg]^{Y}\bigg[\frac{1+r(1+z)^{\xi}}{r} \bigg]  A_1(z) \nonumber \\
&&+ {\rho_{rad0}}(1+z)^{\chi}\Bigg[\frac{-r\beta+1}{-r\beta(1+z)^{\xi}+1}\Bigg]^{Y/\alpha\gamma} A_2(z)\,,
\end{eqnarray}
which with the use of \eqref{RW01} gives the following expression for the Hubble parameter,
\begin{eqnarray}\label{Hubble01-General}
	H^2 &=& H_{0}^2 \Bigg\{(1+z)^{3w_{0}}\Bigg[\frac{-r\beta+1}{-r\beta(1+z)^{\xi}+1}\Bigg]^{Y}\bigg[\frac{1+r(1+z)^{\xi}}{r} \bigg] A_1(z) \Omega_{DM0} \nonumber \\
&&+(1+z)^{\chi} \Bigg[\frac{-r\beta+1}{-r\beta(1+z)^{\xi}+1}\Bigg]^{Y/\alpha\gamma} A_2(z) \Omega_{OM0} \Bigg\}\,.
\end{eqnarray}

By introducing the dimensionless density parameters as,
\begin{eqnarray}\label{DimmensionlessParams01}
	\rho_{c} = \frac{3H_{0}^2}{8\pi G}\,\,, \,
	\Omega_{OM0} = \frac{\rho_{OM0}}{\rho_{c}} \,,\,\,\Omega_{DM0} = \frac{\rho_{DM0}}{\rho_{c}}\,,
\end{eqnarray}
the Hubble parameter reads,
\begin{eqnarray}\label{Hubble-Mz}
	H^2 = H_{0}^2 \Omega_{DM0}M(z),
\end{eqnarray}
where
\begin{eqnarray}\label{MofZ}
	M(z) &=& \Bigg\{(1+z)^{3w_{0}}\Bigg[\frac{-r\beta+1}{-r\beta(1+z)^{\xi}+1}\Bigg]^{Y}\bigg[\frac{1+r(1+z)^{\xi}}{r} \bigg]A_1(z) \nonumber \\
&&+(1+z)^{\chi} \Bigg[\frac{-r\beta+1}{-r\beta(1+z)^{\xi}+1}\Bigg]^{Y/\alpha\gamma} A_2(z) \frac{\Omega_{OM0}}{\Omega_{DM0}} \Bigg\}\,.
\end{eqnarray}

Therefore the dimensionless Hubble parameter can be written as follows,
\begin{eqnarray}\label{EofZ-01}
	E^{2}(z) = \Omega_{DM0} M(z)\,.
\end{eqnarray}

Finally, the deceleration parameter, defined as,
\begin{eqnarray}\label{deceleration01}
	q = -\frac{\dot{H}}{H^2}-1\,,
\end{eqnarray}
 takes in this model the following form,
\begin{eqnarray}\label{deceleration01-total}
	q &=& \frac{1}{2} + \frac{(1+z)^{\chi}}{2E^2(z)} \Bigg[\frac{-r\beta+1}{-r\beta(1+z)^{\xi}+1}\Bigg]^{Y/\alpha\gamma} A_2(z)\Omega_{OM0} +  \nonumber \\ &&\frac{3w_{DE}(z)}{2E^2(z)}  \frac{(1+z)^{3w_{0}}}{r}\Bigg[\frac{-r\beta+1}{-r\beta(1+z)^{\xi}+1}\Bigg]^{Y}A_1(z)\Omega_{DM0} +  \nonumber \\ &&\frac{A_1(z)}{2E^2(z)}{(1+z)^{3w_{0}+\xi}\big(\xi+3w_{DE}(z)\big)}\Bigg[\frac{(1-r\beta)^Y}{(1-r\beta(1+z)^\xi)^{Y+1}} \Bigg]\big(\beta + \frac{1}{\alpha}+1 \big) \Omega_{DM0}\,.
\end{eqnarray}

 The square of the classical sound speed can be expressed as,
\begin{eqnarray}\label{SoundSpeed}
c_{S}^{2}=\frac{(-1+{{\omega }_{0}}+{{\omega }_{1}}{{(1+z)}^{\lambda }})+\lambda {{\omega }_{1}}{{(1+z)}^{\lambda -1}}{{A}_{1}}(z)}{{{A}_{1}}(z)(\frac{3{{\omega }_{0}}}{1+z}-\frac{rY\beta \xi {{(1+z)}^{\xi -1}}}{1-r\beta {{(1+z)}^{\xi }}})-\frac{d{{A}_{1}}(z)}{dz}}\,.
\end{eqnarray}


\subsection{Basic equations of the perturbations of a dissipative cosmological model in the presence of particle creation - the general formalism}

For the sake of completeness, we also briefly present here some basic results on the perturbations of the IDM model. In order to investigate the perturbations of a cosmological model, we need to consider a perturbed metric as given by
\citep{Valiviita:2008iv,Li:2013bya,Xu:2013jma}
\begin{equation}\label{Pert01}
g_{\mu \nu}(\vec{x}, \tau)=\bar{g}_{\mu \nu}(\tau)+\delta g_{\mu \nu}(\vec{x}, \tau)\,,
\end{equation}
where
$\bar{g}_{\mu \nu}(\tau)$ stands for background metric, and $\tau$ denotes the conformal time. For the perturbed terms one has
\begin{equation}\label{Pert02}
\delta g_{00}=-2 a^2 A; \quad  \delta g_{0 i}=a^2 \partial_{i} B; \quad  \delta g_{i j}=a^2\left(-2\psi \gamma_{i j}+F_{i j}\right)\,,
\end{equation}
where $A$ is the lapse function, $\partial_{i} B$ stands for the shift vector, and $F_{i j}=E_{i j}+D_{i j} E$ in which $D^{i j}=\left(\partial^{i} \partial^{j}-\frac{1}{3} \gamma^{i j} \nabla^{2}\right)$. Using the Banachiewicz inversion technique, the contravariant version of Eq.\eqref{Pert02}
can be expressed as
\begin{equation}\label{Pert03}
\delta g^{00}=2 a^{-2} A ;~~~\\
\delta g^{0 i}=a^{-2} \partial^{i} B ;~~~\\
\delta g^{i j}=a^{-2}\left(2 \psi \gamma^{i j}-F^{i j}\right)\,.
\end{equation}

Using Eqs.\eqref{Pert01}-\eqref{Pert03}, the different components of the perturbed Christoffel symbols can be obtained as follows,
\begin{equation}\label{Pert04}
\delta \Gamma_{00}^{0}=A^{\prime};\quad \delta \Gamma_{00}^{i}=\frac{a^{\prime}}{a}\partial^iB+\partial^iB^\prime+\partial^iA;\quad \delta \Gamma_{0j}^{i}=-\psi^\prime \gamma_{ij}+\frac{1}{2}D_{ij}E^\prime;\quad \delta \Gamma_{0i}^{0}=\partial_iA+\frac{a{\prime}}{a}\partial_iB\,,
\end{equation}
and
\begin{equation}\label{Pert05}
\begin{gathered}
\delta \Gamma_{i j}^{0}=-2\left(\frac{a^{\prime}}{a}\right) A \gamma_{i j}-\partial_{j} \partial_{i} B-2\left(\frac{a^{\prime}}{a}\right) \psi \gamma_{i j}-\psi^{\prime} \gamma_{i j}+\left(\frac{a^{\prime}}{a}\right) D_{i j} E+\frac{1}{2} D_{i j} E^{\prime};\quad\\
\delta \Gamma_{j k}^{i}=\partial_{j} \psi \delta_{k}^{i}-\partial_{k} \psi \gamma_{j k}-\frac{a^{\prime}}{a} \partial^{i} B \gamma_{j k}
+\frac{1}{2} \partial_{j} D_{k}^{i} E+\frac{1}{2} \partial_{k} D_{j}^{i} E-\frac{1}{2} \partial^{i} D_{j k} E\,.
\end{gathered}
\end{equation}

Here the superscript $^\prime$ denotes  the derivative with respect to the conformal time. Now, we can turn our attention to the general interacting model of dark energy and dark matter, as described, for example, in Eqs.\eqref{1} and \eqref{2}. We would like to point out that in the following the matter term contains both the cold dark matter, and the radiation components, or, alternatively, the cold dark matter and the ordinary baryonic matter constituents. Moreover, $\varphi$ denotes the dark energy component of the Universe. Hence, the total energy-momentum  tensor of the three-components Universe is given by,
\begin{equation}\label{Pert06}
T^{\mu \nu}=T_{\varphi}^{\mu \nu}+T_{M}^{\mu \nu}\,,
\end{equation}
with
\begin{equation}\label{Pert07}
\nabla_{\mu} T_{\varphi}^{\mu \nu}=Q_{\varphi}^{\nu}, \quad \nabla_{\mu} T_{M}^{\mu \nu}=Q_{M}^{\nu}\,,
\end{equation}
where $\nabla _{\mu}$ denotes the covariant derivative with respect to the metric. Here, $Q^{\nu}$s represent the full covariant form of the quantities introduced in Eqs.\eqref{1} and \eqref{2}, respectively. Generally, they can be decomposed as
\be
Q_{i}^{\nu}=\hat{Q}_{i} u^{\nu}+q_{i}^{\nu}, i={\rm DE, \;M}
\ee
where the index $i={\rm DE, \;M}$ corresponds to dark energy, and matter, respectively, while $q^\nu$ is a four vector perpendicular to the four-velocity $u^\nu$. From Eq.~\eqref{Pert07}, and the definition of $Q^{\nu}$, one immediately obtains,
\begin{equation}\label{Pert08}
u_{\nu} \nabla_{\mu} T_{i}^{\mu \nu}=u_{\nu} Q_{i}^{\nu}=\hat{Q}_{i} u_{\nu} u^{\nu}+u_{\nu} q_{i}^{\nu}=-\hat{Q}_{i}\,,i={\rm DE, \; M}.
\end{equation}
To obtain the above equation we have used the normalization condition of the four-velocity $u_\nu u^\nu=-1$.
Then, with the use of Eqs.\eqref{BalanceDE}-\eqref{BalanceOM}, we can reobtain the parameter $Q$ as written down in Eq.\eqref{Q-Pis}. Following these definitions, one can write down the dissipation functions as $\hat{Q}_i=\pm \tilde{\Gamma}_i \rho_\phi$.

The general form of the perturbations of $Q_i$s can be expressed as
\begin{equation}\label{Pert09}
\delta \hat{Q}_{\varphi}=-\tilde{\Gamma}_1 \delta \rho_{\varphi}-\delta \tilde{\Gamma}_1 \rho_{\varphi}, \;\;
\delta \hat{Q}_{M}=\tilde{\Gamma}_2  \delta \rho_{\varphi}+\delta \tilde{\Gamma}_2 \rho_{\varphi}\,,
\end{equation}

To obtain the perturbed conservation equation, we use the extended form of Eq.\eqref{Pert08}, i.e.,
\begin{equation}\label{Pert10}
\delta\left(u_{\nu} \hat{Q}_i^{\nu}\right)=\delta\left(\underbrace{u_{\nu}\partial_{\mu} T_i^{\mu \nu}}_I+\underbrace{u_{\nu}\Gamma_{\mu \lambda}^{\mu} T_i^{\lambda \nu}}_{II}+\underbrace{u_{\nu}\Gamma_{\mu \lambda}^{\mu} T_i^{\mu \lambda}}_{III}\right)\,,
\end{equation}
where the terms $I$ to $III$ in Eq.\eqref{Pert10} are obtained as follows,
\begin{equation}\label{Pert11}
I=\delta\left(u_{\nu} \partial_{\mu} T^{\mu \nu}\right)=\partial_{0} T^{00} \delta u_{0}+u_{0} \partial_{0} \delta T^{00}
=\frac{1}{a}\left(-2 \mathcal{H} \rho A+\rho^{\prime} A+2 \rho A^{\prime}+2 \mathcal{H} \delta \rho-\delta \rho^{\prime}\right)\,,
\end{equation}
\begin{equation}\label{Pert12}
I I=\frac{\mathcal{H}}{a}(\rho A-\delta \rho)-\frac{1}{a}\left(\rho A^{\prime}+r \rho \psi^{\prime}\right)\,
\end{equation}
and
\begin{equation}\label{Pert13}
I I I=\frac{1}{a}\left(-\rho A^{\prime}+r \omega \mathcal{H} \rho A+r P \psi^{\prime}+\mathcal{H} \rho A-\mathcal{H} \delta \rho-r \omega \delta \rho\right)\,,
\end{equation}
where $\mathcal{H} =a^\prime/a$ is the comoving Hubble parameter. By combining Eqs.\eqref{Pert11} - \eqref{Pert13} in the left hand side of Eq.\eqref{Pert10}, one can obtain
\begin{equation}
\delta \rho^{\prime}+3 \mathcal{H} \delta \rho(1+\omega)-3 \rho(1+\omega) \psi^{\prime}=a \hat{Q} A+a \delta \hat{Q}.
\end{equation}

This equation can describe different eras of cosmological evolutions by introducing the relevant expressions of the equation of state, and of the dissipation functions already obtained in the previous  Sections.

\subsubsection{Perturbing the Einstein field equations}

Now we briefly consider the perturbed Einstein gravitational field equations. For the sake of convenience, we switch to the Newtonian gauge, in which the metric takes the {following form \citep{Riotto,Peter},}
\begin{equation}\label{Pert14}
d s^{2}=a^{2}(\tau)\left[-(1+\Psi) d \tau^{2}+(1+2 \Phi) \gamma_{i j} d x^{x} d x^{j}\right]\,.
\end{equation}
 As compared to Eq.\eqref{Pert02}, it turns out that $\psi$ and $A$ are replaced by the functions $\Phi$ and $\Psi$, respectively. Moreover, in order to reduce the number of degrees of freedom of the system, $E$ and $B$ are taken to be equal to zero. Now, the perturbed Einstein field equations read
\begin{equation}\label{Pert15}
\delta G_{\mu \nu}=\delta R_{\mu \nu}-\frac{1}{2} \delta g_{\mu \nu} R-\frac{1}{2} g_{\mu \nu} \delta R=8 \pi G \delta T_{\mu \nu}\,.
\end{equation}

The different components of the perturbed equations, including the terms $0-0$, $0-i$, and $i-i$, can be obtained as follows
\begin{equation}\label{Pert16}
3 \mathcal{H}^{2}+2 \vec{\nabla}^{2} \Phi-8 \mathcal{H} \Phi^{\prime}=8 \pi G g_{0 \mu} T_{0}^{\mu}
=8 \pi G a^{2} \bar{\rho}(1+2 \Phi+\delta\rho)\,,
\end{equation}
\begin{equation}\label{Pert17}
 \Phi^{\prime}+\mathcal{H} \Phi=-4 \pi G a^{2}(\bar{\rho}+\bar{P}) \vec{v}\,,
\end{equation}
and
\begin{equation}\label{Pert18}
2 \mathcal{H}^{\prime}+\mathcal{H}^{2}+\Phi^{\prime \prime}+3 \mathcal{H} \Phi^{\prime}+\left(2 \mathcal{H}^{\prime}+\mathcal{H}^{2}\right) \Phi=-8 \pi a^{2} \bar{P}+4 \pi a^{2} \delta P\,,
\end{equation}
where $\rho$ and $\bar{P}$ are defined in Eq.~\eqref{tmunu}.

\section{Late-time evolution of the three component model: a particular example}\label{ThreeComp02}

In the following we will present a specific example of the three component interacting dark energy-dark matter model. For this particular case, we restrict our analyze to the following expression for the equation of state parameter of the dark energy,
\begin{eqnarray}\label{EOS-02}
	w_{DE} =  w_{0} - 1\,.
\end{eqnarray}
{In obtaining Eq.~(\ref{EOS-02}) we have assumed that in Eq.~(\ref{Ansatz-EoS}) the parameter $w_1$ satisfies the condition $w_1<<\left(w_0-1\right)/(1+z)^{\lambda}$. On the other hand, this approximation can also be interpreted as an approach for further eliminating a dependence on the redshift parameter of the model, thus obtaining a relatively simpler model as the equation of state of the dark energy. Although this model is apparently simple, it also raises some supplementary problems. By reducing the number of free parameters of the model, it becomes more difficult to fulfill the observational constraints of the cosmological evolution.
We would like also to mention that although it is possible to consider much more complicated versions of the ansatz given by Eq.~(\ref{Ansatz-EoS}), in the present study we investigate only its simplest version, in order to see clearly the behaviour of the considered model with respect to the different observational data sets. Of course, a more complicated equation of state may affect the comparison with the observational results, due to the number of the free parameters of the model.
}
\subsection{Cosmological parameters}

By imposing the above constraint on the dark energy, Eq.\eqref{rhoTotal01} reduces to
\begin{eqnarray}\label{rhoTotal02-B}
	\rho_{tot} =\rho_{DM0}(1+z)^{3w_{0}}\Bigg[\frac{-r\beta+1}{-r\beta(1+z)^{\xi}+1}\Bigg]^{Y}\bigg[\frac{1+r(1+z)^{\xi}}{r} \bigg] + {\rho_{OM0}}(1+z)^{\chi}\Bigg[\frac{-r\beta+1}{-r\beta(1+z)^{\xi}+1}\Bigg]^{Y/\alpha \gamma}
\end{eqnarray}
which gives the following expression for the Hubble rates,
\begin{eqnarray}\label{Hubble02-B}
	H^2 = H_{0}^2 \Bigg\{(1+z)^{3w_{0}}\Bigg[\frac{-r\beta+1}{-r\beta(1+z)^{\xi}+1}\Bigg]^{Y}\bigg[\frac{1+r(1+z)^{\xi}}{r} \bigg] \Omega_{DM0} +(1+z)^{\chi} \Bigg[\frac{-r\beta+1}{-r\beta(1+z)^{\xi}+1}\Bigg]^{Y/\alpha\gamma} \Omega_{OM0} \Bigg\}.
\end{eqnarray}

By taking into account \eqref{DimmensionlessParams01}, one can express the Hubble parameter as
\begin{eqnarray}\label{Hubble-mtildez}
	H^2 = H_{0}^2 \Omega_{DM0}\tilde{M}(z),
\end{eqnarray}
where
\begin{eqnarray}\label{Mtildez}
	\tilde{M}(z) = \Bigg\{(1+z)^{3w_{0}}\Bigg[\frac{-r\beta+1}{-r\beta(1+z)^{\xi}+1}\Bigg]^{Y}\bigg[\frac{1+r(1+z)^{\xi}}{r} \bigg] +(1+z)^{\chi} \Bigg[\frac{-r\beta+1}{-r\beta(1+z)^{\xi}+1}\Bigg]^{Y/\alpha\gamma} \frac{\Omega_{OM0}}{\Omega_{DM0}} \Bigg\}.
\end{eqnarray}
Therefore, the dimensionless Hubble parameter can be written as follows,
\begin{eqnarray}\label{EofMtilde}
	E^{2}(z) = \Omega_{DM0} \tilde{M}(z).
\end{eqnarray}

In this limit the deceleration parameter takes the following form,
\begin{eqnarray}\label{Decelaeration022}
	q &=& \frac{1}{2} + \frac{(1+z)^{\chi}}{2E^2(z)} \Bigg[\frac{-r\beta+1}{-r\beta(1+z)^{\xi}+1}\Bigg]^{Y/\alpha\gamma} \Omega_{OM0} + \nonumber \\ &&\frac{3w_{DE}(z)}{2E^2(z)}  \frac{(1+z)^{3w_{0}}}{r}\Bigg[\frac{-r\beta+1}{-r\beta(1+z)^{\xi}+1}\Bigg]^{Y}\Omega_{DM0} +\nonumber \\ &&\frac{1}{2E^2(z)}{(1+z)^{3w_{0}+\xi}\big(\xi+3w_{DE}(z)\big)}\Bigg[\frac{(1-r\beta)^Y}{(-r\beta(1+z)^{\xi}+1)^{Y+1}} \Bigg]\big(\beta + \frac{1}{\alpha}+1 \big) \Omega_{DM0}\,.
\end{eqnarray}

The square of the classical sound speed can be expressed as,
\begin{eqnarray}\label{SoundSpeed02}
c_{S}^{2}=\frac{-1+{{\omega }_{0}}}{\frac{3{{\omega }_{0}}}{1+z}-\frac{rY\beta \xi {{(1+z)}^{\xi -1}}}{1-r\beta {{(1+z)}^{\xi }}}}\,.
\end{eqnarray}

\subsection{Comparison with the observational data}

\begin{figure}[h]
\centering
\includegraphics[width=0.9\columnwidth]{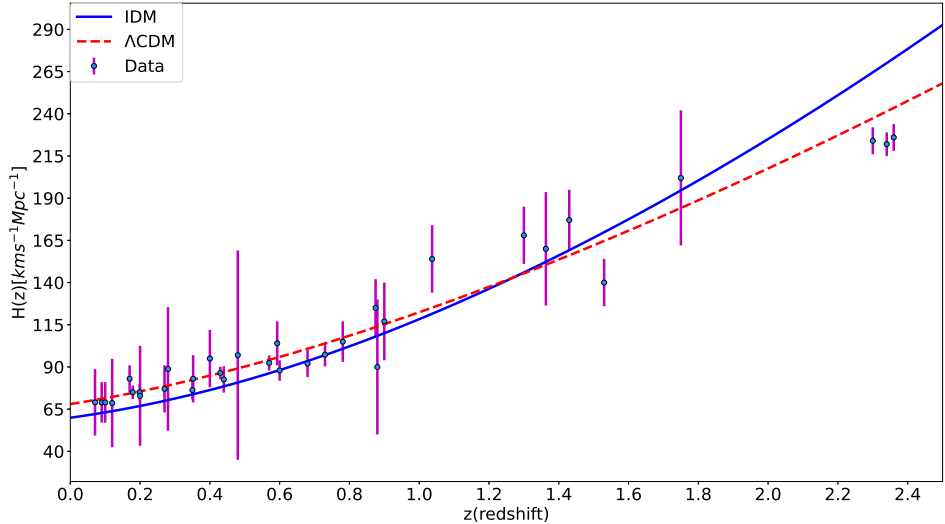}
\caption{In this diagram the blue-solid line shows the evolution of the Hubble parameter versus redshift for the theoretical results based on the interacting IDM model. The predictions of the $\Lambda$CDM model are depicted using a red-dashed line. The blue filled circles with their error bars represent the observational data~\citep{Mostaghel:2016lcd}.}\label{fig:Hubble-rate}
\end{figure}

\begin{figure}[h]
\centering
\includegraphics[width=0.9\columnwidth]{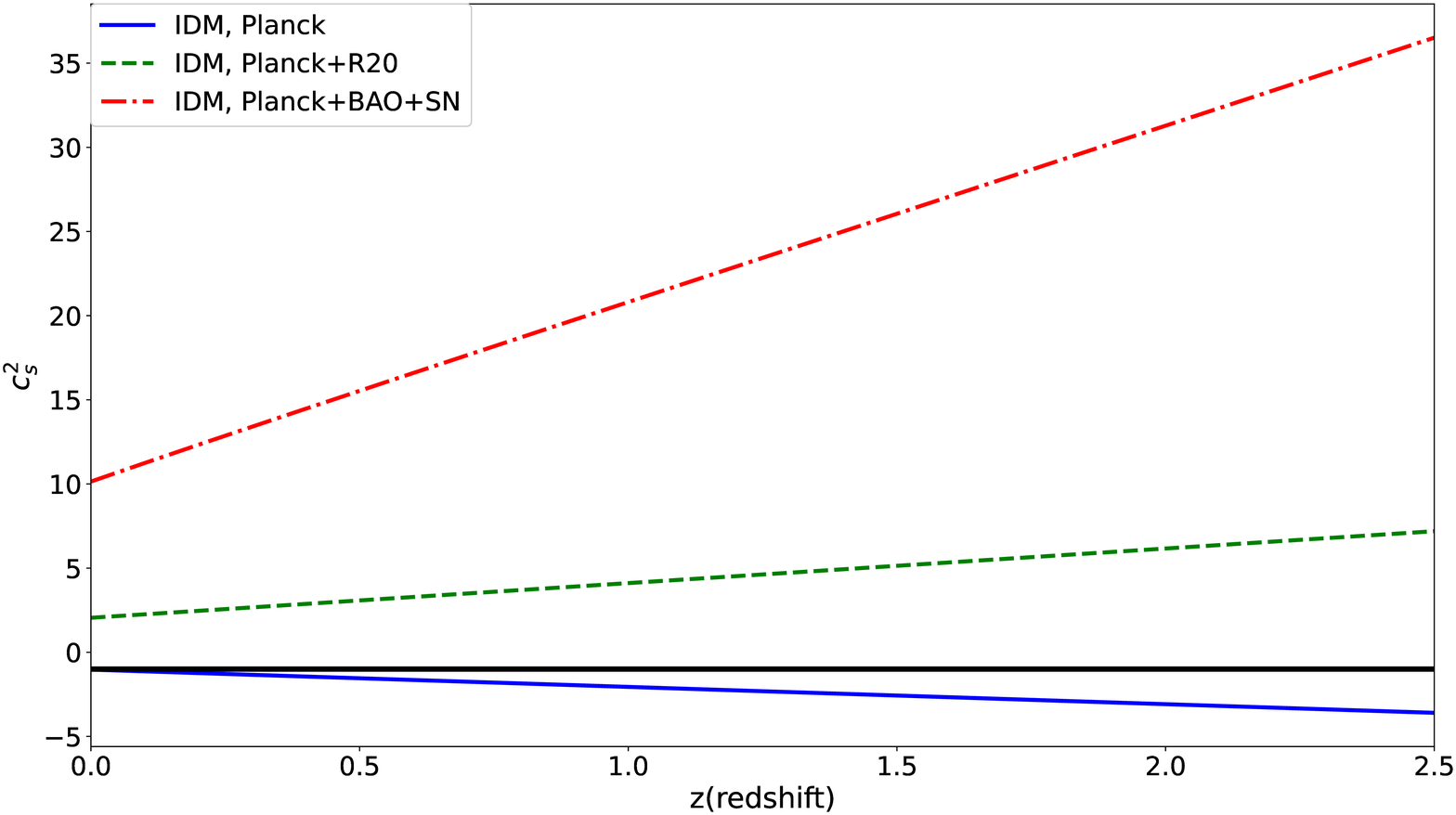}
\caption{{The variation of the speed of sound in the IDM model for various data sets. For the \emph{Planck}+Riess2020 (green dashed line), and for the \emph{Planck}+BAO+SN (red dash-dotted curve) data, the speed of sound shows an acceptable behavior. By considering data from the \emph{Planck} satellite (blue solid curve), the speed of sound has negative values, indicating an instability in model. For the sake of comparison we have also depicted the sound speed  for the $\Lambda$CDM cosmological model (the black solid curve).}}\label{fig:SoundSpeed}
\end{figure}

In the following we will perform a comparison of the predictions of our particular IDM model with the observational data.
As one can see from Fig.~\ref{fig:Hubble-rate}, the Hubble rate of the expansion of the IDM model shows a good concordance with observations. It should be noted that we use the observational data measurements of the expansion rate as presented in ~\citep{Mostaghel:2016lcd}.
 Fig.~\ref{fig:SoundSpeed}

\begin{figure}[h]
\centering
\includegraphics[width=0.9\columnwidth]{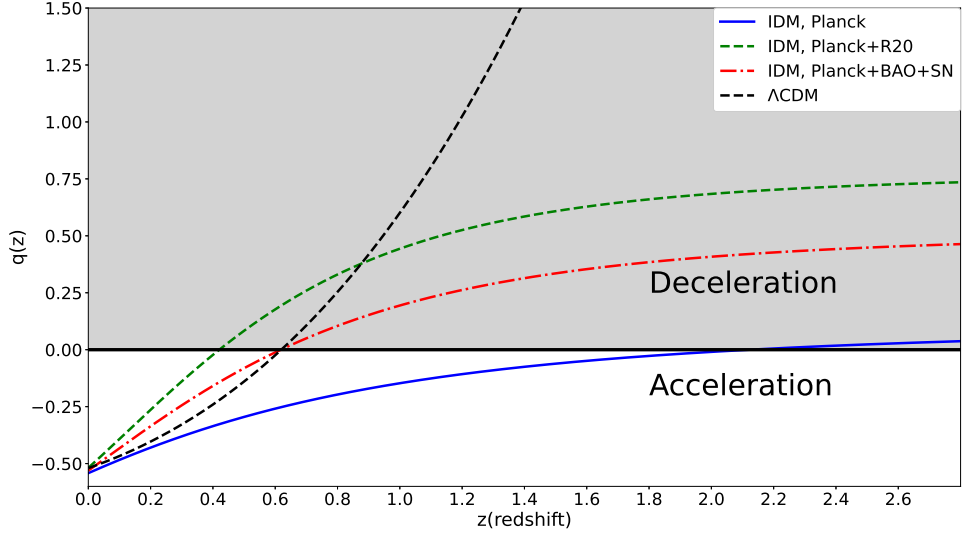}
\caption{{Comparison of the behavior of the deceleration parameter in the IDM model and in the  $\Lambda$CDM model. The black dashed line represents the deceleration parameter in the $\Lambda$CDM model, and the blue solid line shows the best-fit values of the deceleration parameter in the IDM model, based on the \emph{Planck} data. Also, the green dashed and the red dot-dashed lines depict the deceleration parameter for the IDM model based on the combination of the \emph{Planck}+Riess2020 and \emph{Planck}+BAO+SN (Pantheon) data sets.}}
\label{fig:q-z}
\end{figure}

The deceleration parameter of this particular IDM model, given by \eqref{Decelaeration022}, is plotted versus the redshift, by using different data sets, in Fig.~\eqref{fig:q-z}. As one can see from the Figure, the deceleration parameter of the model shows an acceptable behaviour as compared to the $\Lambda$CDM. model predictions.

 The behavior of the squared sound speed versus the redshift $z$ is presented in Fig.~\ref{fig:SoundSpeed}. For the interacting IDM model, different observational data sets have been used to fit the model parameters.

The square of the classical sound speed is an indicator  if the cosmological model is stable or not. To check the stability of this particular three component IDM model, we can analyze the behavior of the speed of sound as given by \eqref{SoundSpeed02}.  From Fig.~\ref{fig:SoundSpeed} it clearly follows that the model is classically stable, and shows a good concordance with respect to the observational data sets.

\subsubsection{The $\mathit{Om}(z)$ diagnostic method}

Another powerful criterion for testing cosmological models is the $\mathit{Om}$ diagnostic method. One can use this technique to determine the behavior of theoretical  models, and to perform comparison with the observational data. The parameter $\mathit{Om}$ is a geometrical diagnostic, which can be defined as  a combination of the Hubble parameter and of the redshift. This parameter can distinguish between various dark energy models, and the $\Lambda \rm{CDM}$ paradigm. To test the three component IDM model we consider a  diagnostic tool $\mathit{Om}(z)$ that is defined as follows (see \citep{Sahni:2008xx}),
\begin{equation}
\mathit{Om}(z) \equiv \frac{E^2(z) -1}{(1+z)^3 -1}\,,
\end{equation}
where
\begin{equation}
E^2(z) = \frac{H^2(z)}{H_0^2}\,.
\end{equation}

\begin{figure}[h]
\centering
\includegraphics[width=0.9\columnwidth]{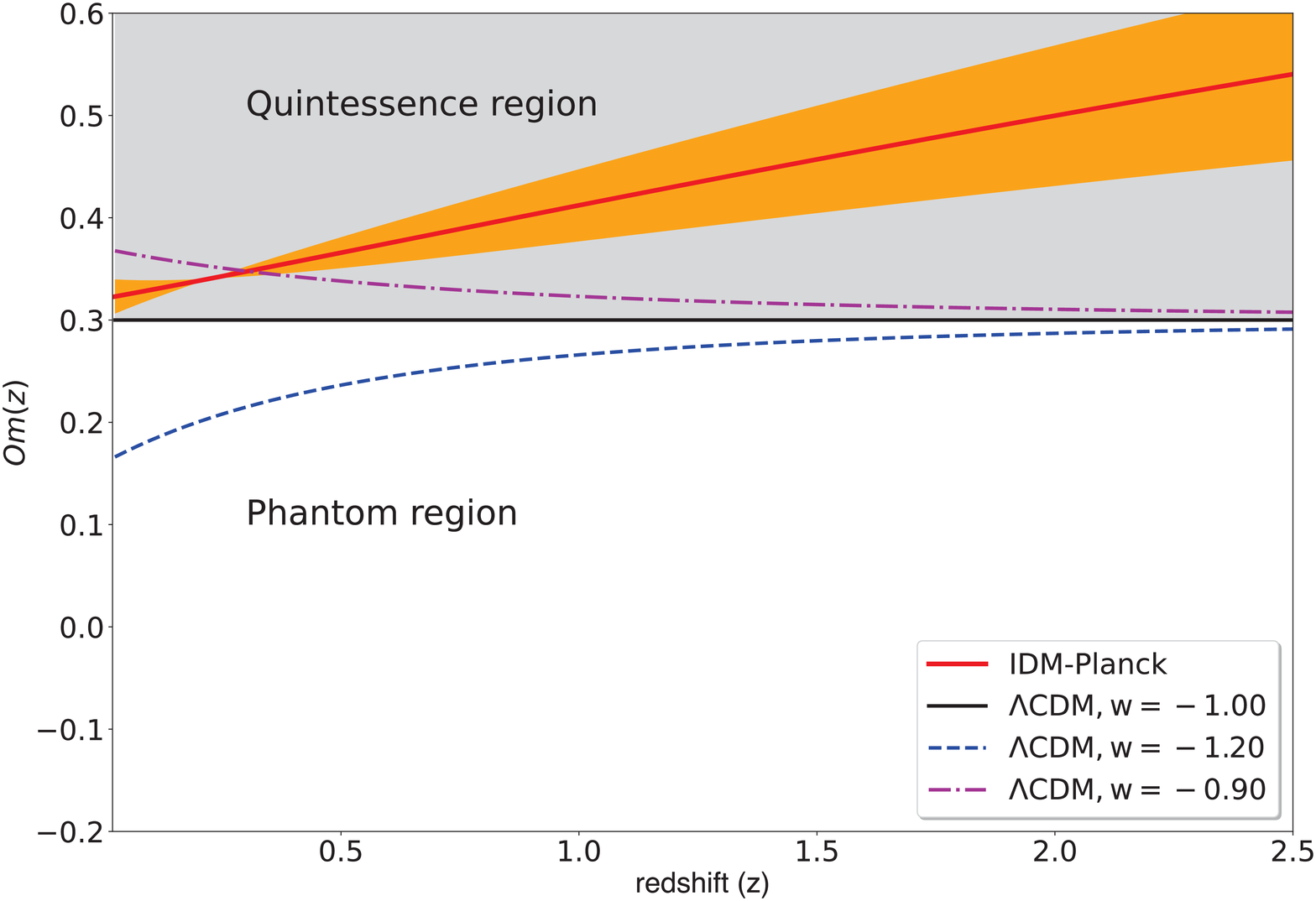}
\caption{{The $\mathit{Om}(z)$ diagnostic: the solid black curve represents $\mathit{Om}(z)$ for the $\Lambda$CDM model. The dashed and dash-dotted curves represent $\mathit{Om}(z)$ for $w=-1.20$ and $w=-0.90$, respectively. The red solid curve above the 1$\sigma$ confidence level (orange shade) shows the best fit values of $\mathit{Om}(z)$ for the IDM model, determined by using the \emph{Planck} data.}}
\label{fig:om-z}
\end{figure}

In Fig.~\ref{fig:om-z} one can see the behavior of $\mathit{Om}(z)$ for the IDM model.
For the $\Lambda \rm{CDM}$ model $\mathit{Om}(z)=\Omega_m$, while for the other dark energy models,  $\mathit{Om}(z)$ depends on the redshift \citep{Shafieloo:2014ypa}. Phantom like dark energy corresponds to the positive slope of $\mathit{Om}(z)$, whereas the negative slope means that dark energy behaves like Quintessence \citep{Shahalam:2015lra}. In addition, $Om(z)$ depends upon the higher derivatives of the luminosity distance, and, therefore, as compared with $w(z)$ and the deceleration parameter $q(z)$, it is less sensitive to the observational errors \citep{Sahni:2008xx}.

In the next Section, by employing different observational data sets,  we will carefully address the behavior of these parameters, and we will try to constrain the free parameters of the IDM model as well.


\section{Methodology and Observational Data}\label{MOD}

In this Section, we use a combination of recent early universe and late time cosmological measurements to constrain the three component IDM model as follows:

\begin{itemize}
\item{ \bf Cosmic Microwave Background:} We consider the CMB temperature, polarization, and lensing reconstruction angular power spectra as measured by 2018 Planck legacy release ~\citep{Aghanim:2018eyx,Aghanim:2019ame}. We denote them as "Planck," and they include the CMB temperature and polarization data (TT, TE, EE+lowE; where the low-multipole polarization is obtained from the High-Frequency Instrument, HFI).

\item {\bf Hubble Space Telescope (HST):} We also use the most recent constraints inferred via the Cepheid-calibrated SNIa distance ladder from SH0ES, $H_0 = 73.2 \pm 1.3$ km/s/Mpc~\citep{Riess:2020fzl}, and we denote the corresponding data as "R20".

\item {\bf BAO:} As a probe of the (relative) cosmic expansion history at low redshifts, we consider baryon acoustic oscillation (BAO) data from the SDSS DR7 main galaxy sample~\citep{Ross:2014qpa} at $z = 0.15$, the 6dF galaxy redshift survey~\citep{Beutler:2011hx} at $z = 0.106$, and from the SDSS BOSS DR12~\citep{BOSS:2016wmc,C1} LOWZ and CMASS galaxy samples at $z = 0.38$, $0.51$, and $0.61$.  The BAO data do not provide absolute distances to these redshifts, but rather relative distances normalized to the sound horizon at the end of the baryon drag epoch. To be conservative, we do not consider redshift-space distortion or full-shape galaxy power spectrum data in this work.

\item {\bf Supernovae data:} The recent Pantheon sample of SNe Ia, including 1048 data points, with $0.01 < z < 2.26$ ~\citep{Scolnic:2017caz}, has been applied to constrain numerous cosmological models ~\citep{Wang:2018ahw}, and it is also employed in our cosmological analysis.

\end{itemize}

In our cosmological analysis, we perform Markov Chain Monte Carlo (MCMC) calculations with a modified version of  the publicly available code \textsc{CosmoMC} \citep{Lewis:2002ah}. We use a convergence criterion that obeys $R-1 < 0.01$, where the Gelman-Rubin $R$-statistics \citep{Gelman92} is the variance of chain means divided by the mean of chain variances.\\

In our analysis we consider a 11 parameters model, with 6 of them being the $\Lambda$CDM model parameters, plus 5 extra parameters. The assumptions for the priors of the parameters are listed in Table~\ref{tabpriors}.
\begin{table}[tb]
	\begin{center}		\resizebox{0.33\textwidth}{!}{
\begin{tabular}{||c||c||}
\hline
\hline
Parameter & Prior\\
\hline
$\alpha $ & $[0.1, 10]$ \\
\hline
$\beta$ & $[10,20]$ \\
\hline
$\gamma$ & $[0.0001, 10]$ \\
\hline
$\xi$ & $[0.5,6.0]$ \\
\hline
$w_0$ & $[-0.4,3.0]$ \\
\hline
 $\Omega_{\mathrm c}h^2$ & $[0.001, 0.99]$\\
 \hline
 $\Omega_{\mathrm b}h^2$ & $[0.005, 0.1]$\\
 \hline
 $\ln{(10^{10} A_{\mathrm s})}$ & $[1.61, 3.91]$\\
 \hline
$n_{\mathrm s}$ & $[0.8, 1.2]$\\
\hline
 $100 \Theta_{\rm {MC}}$ & $ [0.5, 10.0]$\\
 \hline
$\tau$ & $[0.01, 0.8]$ \\
\hline
 $k_{\rm pivot}$ & $0.05$ \\
 \hline
\hline
\end{tabular}
}
\caption{Flat priors for the variation of  the cosmological parameters for the three components IDM model. In the Table $\Omega _ch^2$ is the cold dark matter density today, $\Omega _bh^2$ is the baryon density today, $100 \Theta_{\rm {MC}}$ is the 100 $\times$ approximation to $r_*/D_{\Lambda}$ (CosmoMC), $\ln{(10^{10} A_{\mathrm s})}$ is the Log power of the primordial curvature perturbations, $n_s$ is the Scalar spectrum power-law index, and $\tau$ is the Thomson scattering optical depth due to reionization, respectively.} \label{tabpriors}
\end{center}
\end{table}


\begin{table*}[tb]
	\begin{center}		\resizebox{0.8\textwidth}{!}{
		\begin{tabular}{|| c  ||c||c|| c ||  }
				\hline
				\hline
				Parameters &   Planck &Planck+R20 & Planck+BAO+Pantheon  \\ \hline

				$\alpha$ &  $> 4.14$ & $> 4.25$ & $> 4.20$ \\
\hline
             	$\beta$ &  $> 13.7$ & $> 13.7$  & $> 13.8$\\
             \hline
             	$\gamma$ &  $< 5.92$ & $< 6.00$ & $< 6.10$ \\
             \hline
             	$\xi$ &  $2.24\pm 0.66$ & $3.54\pm 0.17$ & $3.04\pm 0.11$\\
             \hline
             	$w_0$ &  $0.25\pm 0.23$ & $-0.184\pm 0.066$ & $-0.022\pm 0.050$ \\
             \hline
				$\Omega_b h^2$ &  $0.02209\pm 0.00025$ & $0.02215\pm 0.00025$ & $0.02218\pm 0.00023$\\
\hline
				$\Omega_c h^2$ &  $0.1209\pm 0.0021$ & $0.1198\pm 0.0021$  & $0.1191\pm 0.0016$\\
\hline
				
				$\Omega_m$  &  $0.413^{+0.070}_{-0.10}$& $0.2706^{+0.0098}_{-0.011}$ & $0.3076\pm 0.0079$\\
\hline
				
				$H_0$ &  $59.9^{+5.3}_{-7.2}$ & $72.6\pm 1.3 $ & $67.95\pm 0.82$ \\
\hline
				
				$100\theta_{MC} $  & $1.04074\pm 0.00048$ & $1.04086\pm 0.00048$  & $1.04095\pm 0.00045$ \\
\hline
				${\rm{ln}}(10^{10} A_s) $  &  $3.042\pm 0.015$ & $3.040\pm 0.015$ & $3.040\pm 0.015$ \\
\hline
				$n_s $ & $ 0.9620\pm 0.0058$  & $0.9651\pm 0.0057$   & $0.9663\pm 0.0049$\\
\hline
				$\tau$ &  $0.0526\pm 0.0059$ & $0.0533\pm 0.0059$  & $0.0541\pm 0.0058$\\
\hline
				$\chi^2_{\rm CMB}$& $630.8\,({\nu\rm{:}\,7.5}) $ & $633.3\,({\nu\rm{:}\,7.2}) $  & $631.0\,({\nu\rm{:}\,6.4}$\\
\hline
				$\chi^2_{\rm H073p20}  $  & $--$ & $1.2\,({\nu\rm{:}\,1.6})   $ & $ --$ \\
\hline
				$\chi^2_{\rm BAO}$ & $--$ & $--$& $6.3\,({\nu\rm{:}\,0.9})$ \\
\hline
				$\chi^2_{\rm JLA}$ & $--$ & $--$ & $1035.5\,({\nu\rm{:}\,0.5})$ \\

				\hline
				\hline
				
			\end{tabular}
	}	
	\end{center}
	\caption{$68\%$ CL constraints on the IDM model, considering \emph{Planck} data with Riess 2020, BAO,  and also supernovae Pantheon data. In the Table $\Omega_bh^2$, $\Omega_ch^2$, and $\Omega_m$ are the baryon density today,  the dark matter density today, and the matter density today divided by the critical density, respectively, while $H_0$ denotes the  Hubble parameter at the present time. Moreover, $100 \Theta_{\rm {MC}}$ denotes the 100 $\times$ approximation to $r_*/D_{\Lambda}$ (CosmoMC), $\ln{(10^{10} A_{\mathrm s})}$ is the Log power of the primordial curvature perturbations, $n_s$ is the Scalar spectrum power-law index, and $\tau$ is the Thomson scattering optical depth due to reionization, respectively. The subscript  $JLA$ in $\chi^2_{\rm JLA}$ refers to the Joint Lightcurve Analysis.}
	\label{table1}
\end{table*}

\subsection{Observational constraints} \label{ThreeCompObservational}

In the following we will perform a systematic investigation, and comparison, of the three component IDM model with the observational data.

\subsubsection{Brief review of the Cosmic Microwave Background Radiation}

To better understand the physics of the problem, here we will first present a brief overview of the Cosmic Background Radiation, and of its temperature anisotropies \citep{EcyclopediaCosmo}. Although the Cosmic Background Radiation has one of the most complete spectra of black-body radiation, its temperature distribution has a very small perturbation of the order of $\Delta T/T\sim 10^{-5}$. This value, though seemingly insignificant, provides a very rich physics for understanding the behaviour of the Universe from the early phases of its formation.

As a first application, this anisotropy can be used to measure distances on a cosmic scale. It is conceivable that the contents of OMr in the early Universe, including baryons and photons, were coupled to each other, and oscillated in phase with each other. This configuration, and its evolution, definitely creates hot and cold spots in the background radiation map, that is, in the  CMB map, making it possible to estimate the curvature of the Universe by measuring the angular scales of the temperature anisotropy. The assumption that the nature of these anisotropies is statistical does not lead to many theoretical or computational complications. Hence, the statistical study of this phenomenon can be done with the help of the angular power spectrum of the temperature anisotropy \citep{Aghanim:2019ame}.

For mathematical convenience, it is better to express the temperature anisotropy in terms of the spherical harmonics,
\begin{equation}\label{Harmo-Sph}
\frac{\Delta T}{T}(\hat{\boldsymbol{n}})=\sum_{\ell, m} a_{\ell}^{m} Y_{\ell m}(\hat{\boldsymbol{n}})\,,
\end{equation}
where $ a_{\ell}^{m}$ are the harmonic coefficients. It can also be noted that ${\ell}=0$ refers to the monopole anisotropy, ${\ell}=1$  refers to the dipole anisotropy, and ${\ell}\geq2$  stands for multipolar anisotropies. Also, to define the correlation between two points in the sky that are separated by the angle $\theta$, we can use the definition of the angular autocorrelation function given by
\begin{equation}\label{Authocorrel}
C(\theta)=\left\langle\frac{\Delta T}{T}\left(\hat{n}_{1}\right) \frac{\Delta T}{T}\left(\hat{n}_{2}\right)\right\rangle\\
 =\frac{1}{4 \pi} \sum_{\ell} \sum_{m=-\ell}^{m=+\ell}\left|a_{\ell}^{m}\right|^{2} \mathcal{P}_{\ell}(\cos \theta) \\
=\frac{1}{4 \pi} \sum_{\ell}(2 \ell+1) \mathcal{C}_{\ell} \mathcal{P}_{\ell}(\cos \theta)\,,
\end{equation}
where $\mathcal{P}_{\ell}$ stands for the Legendre polynomials,  and $\mathcal{C}_{\ell}$  is the expectation value, defined as $\left\langle\left|a_{\ell}^{m}\right|^{2}\right\rangle$ \citep{Aghanim:2019ame,EcyclopediaCosmo}.  Interestingly, one can use the definition of this expectation value to measure the broadband of the power spectrum in terms of the $\log \ell $ as
\begin{equation}\label{Broadband}
\mathcal{D}_{\ell}^{T T}=\frac{\ell(\ell+1)}{2 \pi} \mathcal{C}_{\ell}\,.
\end{equation}

\subsection{Observational constraints of the IDM model}

\begin{center}
\begin{figure}[htbp]
\center
\includegraphics[width=0.9\columnwidth]{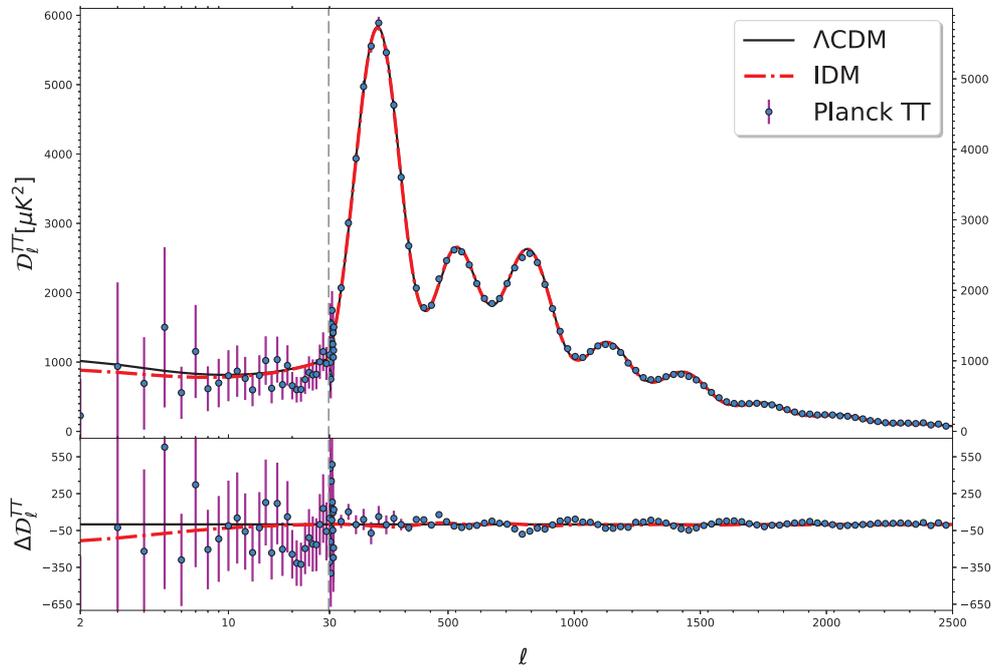}
\caption{{\textit{Upper panel:} The power spectrum for the CMB temperature anisotropies for the IDM model is shown comparatively with the \emph{Planck} 2018 data, and with the standard $\Lambda$CDM model. \textit{Lower panel}: Residuals of the power spectra with respect to the standard-$\Lambda$CDM model.}}
\label{fig:powerspectrum}
\end{figure}
\end{center}

The evolution of the power spectrum for the CMB anisotropies in the IDM model is depicted in Fig.~\ref{fig:powerspectrum}. In the Figure the best fits of both the IDM model and the $\Lambda$CDM model are presented as well. The first peak appears at $\sim \ell =200 (\pi/\theta\sim1 \text{degree})$.
Another parameter that is important in power spectrum measurements is the residual. This parameter deals with the relationship between $\mathcal{C}_{\ell}$,  and the changes in  the parameters of the model. It is usually defined as (see \citep{Planck:2016tof,C4}),
\begin{equation}
\Delta \mathcal{D}_{\ell}^{TT}=\partial \mathcal{C}_{\ell} / \partial p_{i}\,.
\end{equation}

\begin{figure}[htbp]
\centering
\includegraphics[width=0.8\columnwidth]{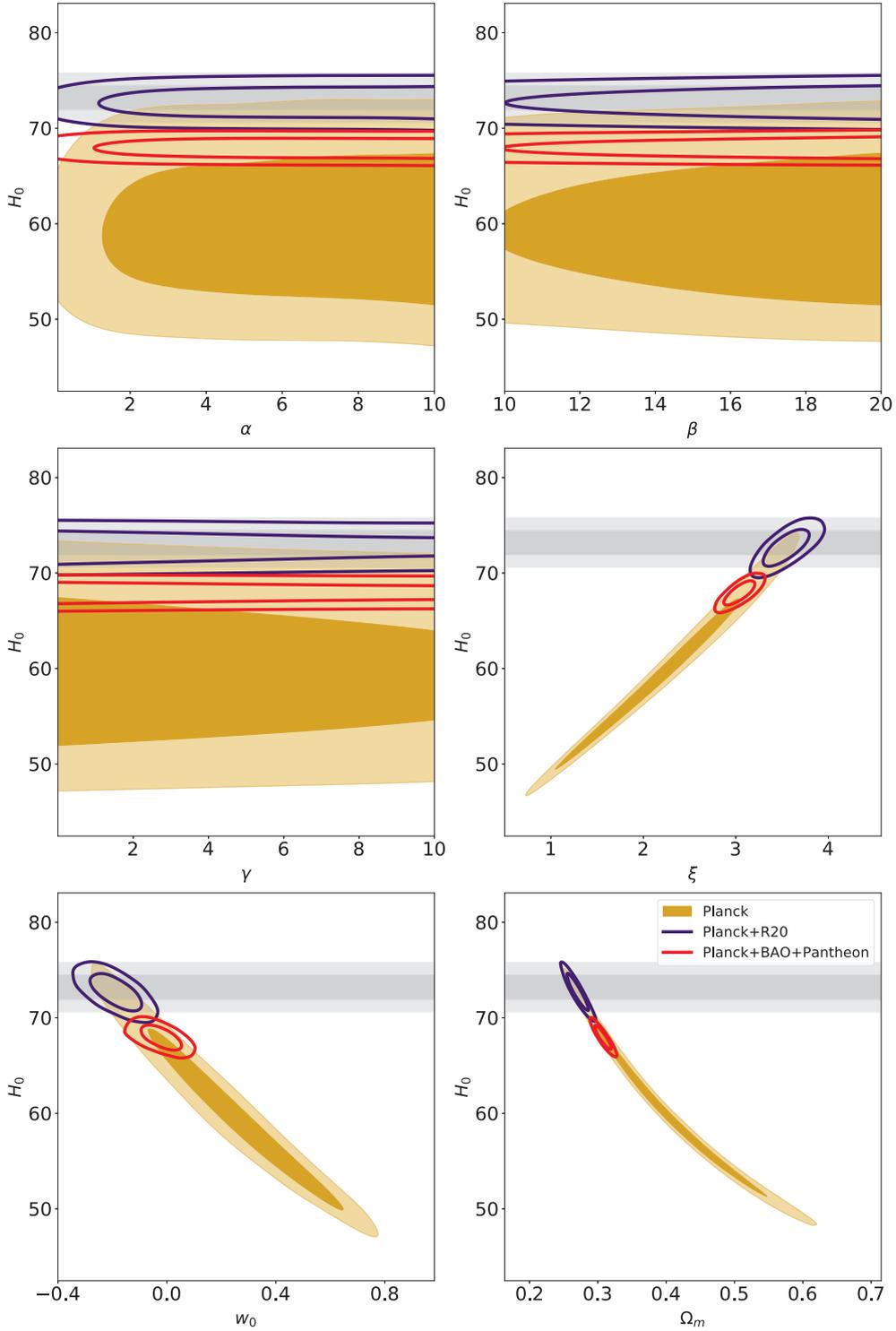}
\caption{{These two dimensional diagrams show the contour plots for the parameter $H_0$ versus different parameters of the IDM model. Confidence levels for $1\sigma$ and $2\sigma$ are obtained based on \emph{Planck} data, Riess2020, BAO and Pantheon supernovae data. The gray shaded areas show the best-fits and the error-bars of Riess et.al measurement of $H_0$,  respectively~\citep{Riess:2020fzl}.}}\label{fig:2d-h0}
\end{figure}

Also, one of the methods for checking the correctness of a theoretical model is to examine the free parameters appearing in the model under the observational constraints, and to present the related contours. For this purpose, we have carefully studied the effect of changes in various IDM model parameters on the behavior of the Hubble parameter. The results of these investigations are shown in Fig.~\ref{fig:2d-h0}. In these counter plots the confidence levels for $1\sigma$ and $2\sigma$ are obtained by using the \emph{Planck} data, Riess2020, BAO and Pantheon supernovae data, respectively.

\begin{figure}[htbp]
\centering
\includegraphics[width=0.8\columnwidth]{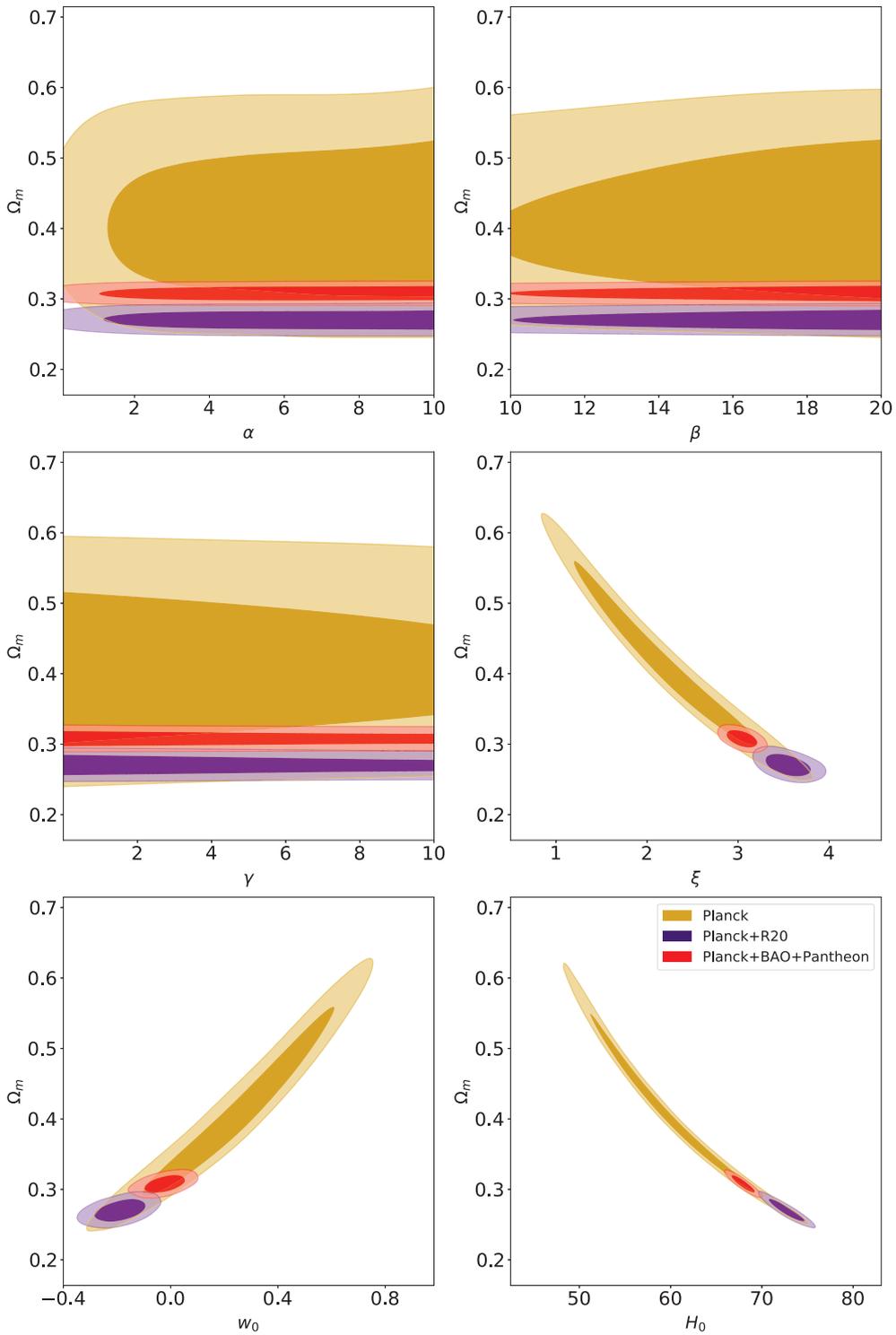}
\caption{{Contour plots for $1\sigma$ and $2 \sigma$ confidence levels for the parameter $\Omega_m$ versus different parameters of the IDM model. The confidence levels for $1\sigma$ and $2\sigma$ are plotted  based on the \emph{Planck} data, Riess2020, BAO and Pantheon supernovae data sets.}}\label{fig:2d-omegam}
\end{figure}

It should be noted that the gray shaded areas represent he best-fits and error-bars of Riess et.al measurements of the $H_0$ values, respectively~\citep{Riess:2020fzl}. Another parameter that plays an important role in the study of cosmic evolution is the mass density parameter, whose behavior is plotted in terms of various parameters, as well as of the Hubble parameter, in Fig.~\ref{fig:2d-omegam}.

\begin{figure}[htbp]
\centering
\includegraphics[width=0.8\columnwidth]{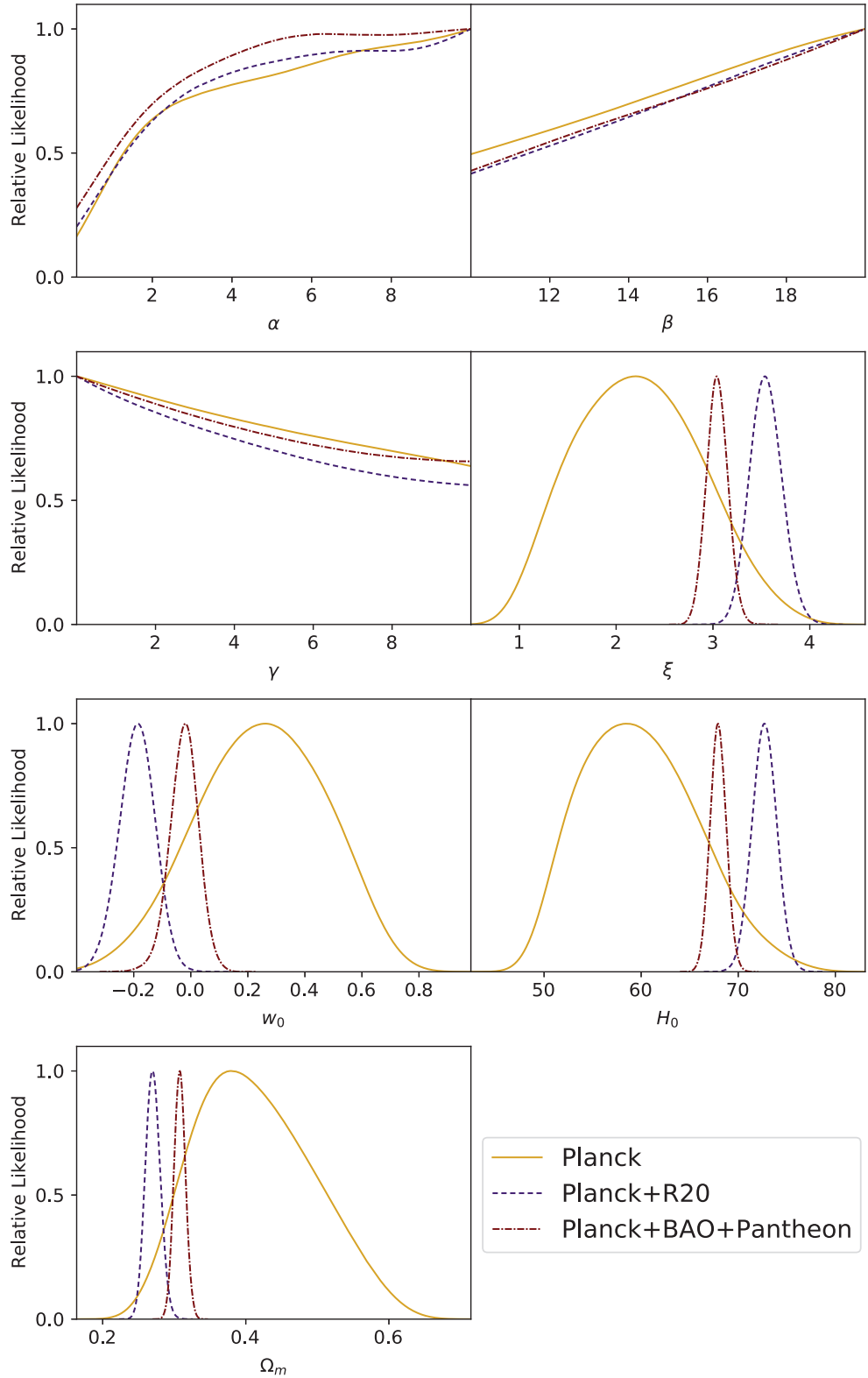}
\caption{{Relative likelihoods for different parameters of the IDM model, based on different combinations of data sets. The diagrams for the parameters $\alpha$, $\beta$ and $\gamma$ indicate that the available data are not sufficient, or accurate enough, to constrain the parameters as a Gaussian likelihood. For the $H_0$ parameter, and for the \emph{Planck} data, the sequence of the likelihood diagram, i.e., $2 \sigma$, is close to the results of the Riess2020 data. However, at the $1 \sigma$ level, and at its best-fit, the values predicted by the IDM model have some differences as compared to the R20 predictions.}}
\label{fig:1d-likes}
\end{figure}

\begin{center}
\begin{figure}[htbp]
\centering
\includegraphics[width=0.87\columnwidth]{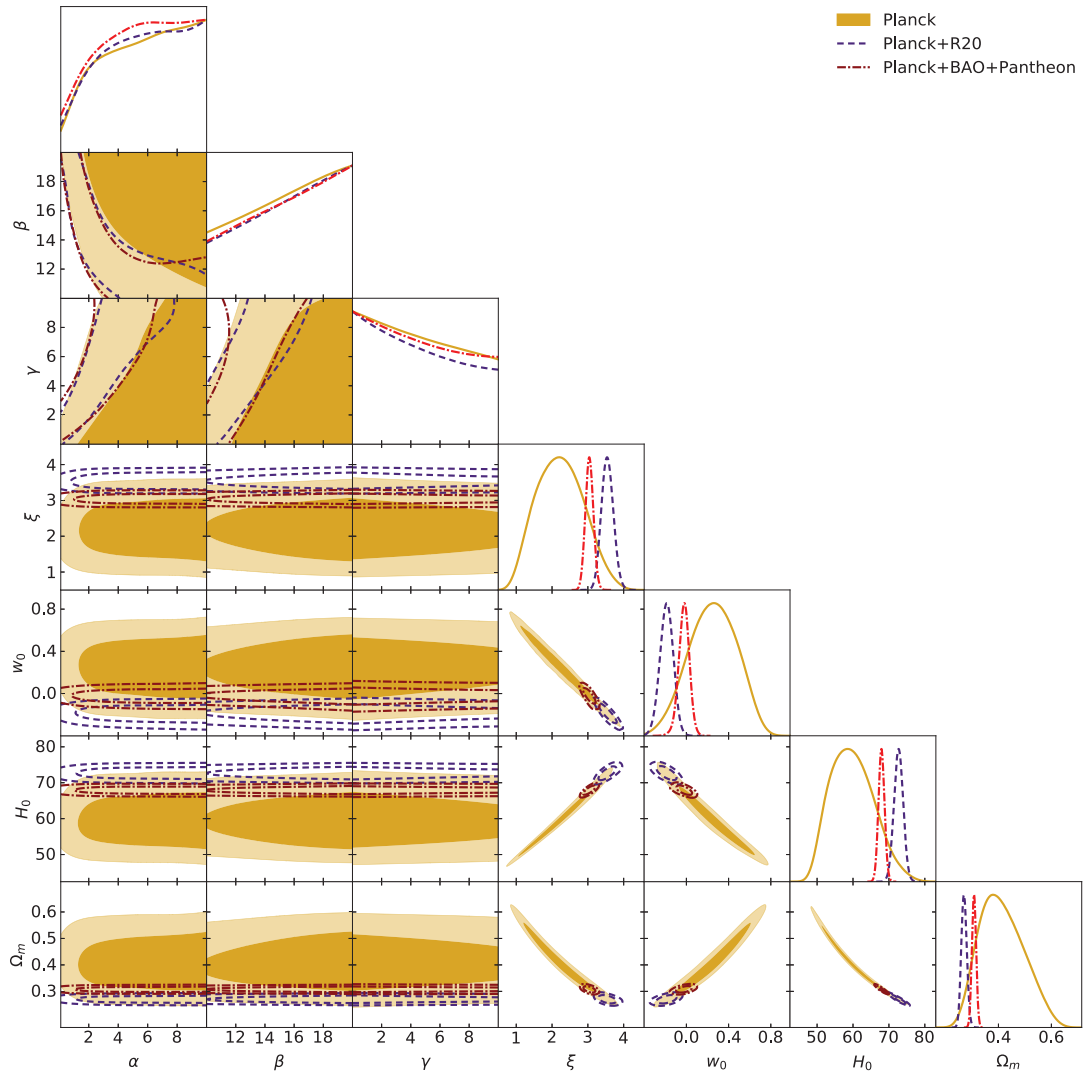}
\caption{{The triangular plot, including two-dimensional contour plots, and one-dimensional relative likelihoods, is presented for different IDM model parameters. The filled contours  represent the \emph{Planck} data alone. The dashed and the dot-dashed curves represent the \emph{Planck}+Riess2020 and \emph{Planck}+BAO+SN data.}}
\label{fig:tri_plot}
\end{figure}
\end{center}

Another diagram that plays a crucial  role in the behavioral study of model parameters is the one-dimensional  relative likelihood diagram, represented in Fig.~\ref{fig:1d-likes}. In the graphs presented in the Figure, depending on which category of data we have used, we can obtain an estimate of the model parameters, and their best-fits. However, more exactly, these diagrams for the parameters $\alpha$, $\beta$ and $\gamma$ indicate that the available data are not sufficient or accurate enough to constrain the model parameters as a Gaussian likelihood. For the $H_0$ parameter and for the \emph{Planck} data, although in the tail of the  diagram, i.e., at the $2 \sigma$ level, there is an overlap with the results of the Riess2020,  at the $1 \sigma$, the values predicted by the IDM model are in conflict with the R20 predictions. In addition to these results, to obtain a better comparison of the results with the observations,  and to see how the various parameters change in the model, it is common to use triangular diagrams that include one-dimensional and two-dimensional diagrams simultaneously. For this purpose, we present such a diagram in Fig.~\ref{fig:tri_plot}.

\newpage







\section{Concluding remarks}\label{Conclustions}

In the present work we have investigated in detail a three-component model of the Universe, with the total energy content given by $\left(\rho_{DE},\rho_{DM},\rho_{OM}\right)$, by also allowing for the possibility of an interaction between the constituents. The interaction leads to the nonconservation of the total energy–momentum of the cosmological matter, and it represents at the same time an interesting, and intriguing aspect of the cosmological dynamics. We have investigated the implications and  the significance of the matter creation and decay processes in the three component Universe by adopting the theoretical approach of the thermodynamics of irreversible processes in open systems, as first introduced and extended in \citep{Prig,Cal}. Our basic assumption is that the interaction between the components of the Universe leads to transfer of energy from the dark energy to various forms of matter. As a result, at the cosmological level a temporal variation in the particle number density does occur. Thus, through matter production, dark energy may also act as a source of entropy, and even internal energy for ordinary matter, respectively. Matter creation takes place in an expanding Universe, and hence one can establish a strong correlation between matter creation and the overall cosmological
dynamics. Large levels of matter creation may determine the transition from deceleration to acceleration in the expanding Universe.

As a result of the decay of the dark energy, dark matter and ordinary matter are produced at rates that are proportional to the energy density of the dark energy $\rho_{DE}$. Systems in which particle creation and decay takes place can be naturally described by using the thermodynamics of open systems, as initially proposed in \citep{Prig}, and further developed in \citep{Cal}. The basic thermodynamic quantity that describes matter decay or formation is the creation pressure, an effective quantity that is proportional to the energy density of the components, but also depends, in a cosmological context, by the Hubble function. In order to close the model we need to impose some functional relations between the particle creation and decay rates, which we have formulated as $\Pi_{DE}=\alpha \beta\Pi_{OM}$, and $\Pi_{DM}=\alpha \Pi_{OM}$, indicating that decay rate of dark energy and the creation rate of dark matter are ultimately proportional to the ordinary matter creation rate. We have also assumed a close proportionality relation between the energy density of the ordinary and dark matter, respectively. Hence, the present model describes the cosmological dynamics, and especially the most recent observations, via matter creation processes. In order to obtain such a description, two supplementary assumptions are also necessary, namely the scaling of the ratio of the dark matter and dark energy with respect to the redshift, which we take as $\rho_{DM}/\rho_{DE}\sim (1+z)^\xi$, and a functional analytic form of the equation of state of the dark energy, $w_{DE}$, which can also be parameterized in terms of the redshift. Once the model parameters are fixed, a full comparison with the observational data can be performed.

We have also considered a specific theoretical model, in which the parameter of the dark energy equation of state was assumed to have the simple form $w_{DE}=w_0-1$, with $w_0$ a constant. Such an equation of state for the  dark energy can describe the late time evolution of the Universe, and for $w_0=0$ it reduces dark energy to a cosmological constant. Non-zero values of $w_0$ can describe the possible deviations of the dark energy from a pure $\Lambda$. The comparison of the behavior of the Hubble function with the observations shows an acceptable concordance with the observational data, and with the predictions of the $\Lambda$CDM model. Moreover, important differences do appear in the behavior of the deceleration parameter, but with the differences essentially depending on the observational data sets used for fitting the model parameters. If one considers the Planck data alone, the IDM model predicting a more rapid expansion of the Universe, as compared to the $\Lambda$CDM model. If one uses for fitting the Planck+Bao+SN data sets, one obtains a qualitatively similar description of $q$ as given by the $\Lambda$CDM model. For the same set of data Planck+Bao+SN the model is stable in the sense that the effective speed of sound turns out to be positive for all redshifts.

In fact, particle production can easily explain the accelerated de Sitter expansion of the Universe without assuming the presence of dark energy. The reasoning is as follows: due to the expansion of the Universe, the matter density would decrease as $\rho(t)=\rho_0\exp(-3H_0t)$, where $\rho_0$ is the present matter density of the Universe. Thus, the time variation of the density is given by $\dot{\rho}(t)=-3H_0\rho_0\exp(-3H_0t)$. We estimate this relation at the present time, and we also take $\rho_0=3H_0^2/8\pi G$. Then we obtain $\left.\dot{\rho}(t)\right|_{t=0}=-9H_0^3/8\pi G$. Therefore, the particle creation rate necessary to keep the matter density constant is
\be
\Gamma|_{t=0}=- \left.\dot{\rho}(t)\right|_{t=0}=\frac{9H_0^3}{8\pi G}.
\ee

With this creation rate the matter density is constant, and the evolution is de Sitter, with $3H^2=8\pi G \rho_{OM}={\rm constant}$. Now let's estimate the matter creation rate $\Gamma $ numerically. Taking $H_0=2.2\times 10^{-18}\; {\rm s}^{-1}$ (Planck data), we obtain first $\Gamma=5.71\times 10^{-47}\; {\rm g/cm ^3}/s$. We convert cm to km and seconds to years, thus obtaining $\Gamma =1.8\times 10^{-24}\;{\rm  g/km}^3/{\rm year}$. This means that on a cosmological scale the creation of a single proton in one km$^3$ in one year, or 160 protons in a century,  can compensate the density decrease of matter due to the de Sitter expansion. It is obvious that such a small amount of created matter is beyond the observational or experimental reach.

An interesting and important implication of the cosmological matter production is its relation with the problem of the arrow of time. This problem consists in obtaining a physical mechanism that could generate a linear evolution of time, which would allow us to differentiate between the past and the future of the
Universe. We can in fact introduce two different arrows of time. The first arrow of time is
created thermodynamically,  and it is fully determined by increase of the entropy of
the Universe. But one can also consider the cosmological arrow of time, determined by the
direction of the expansion of the Universe. Particle creation via the decay of the dark energy determines an asymmetry in the
temporal expansion of the Universe, thus allowing the introduction of a thermodynamical arrow of time, which is fully
determined by the matter production processes. In the cosmological scenario considered in the
present work, it turns out that the thermodynamical arrow of time, determined by matter creation, is identical with the cosmological
one, determined by the temporal evolution of the Universe. Hence, both arrows of time points towards a single time arrow
describing global evolution.

{In the present work we have also introduced, at a purely theoretical level, and in their general formulation, the basic equations describing the perturbations  of the IDM model. The evolution of the cosmological perturbations play a key role in the study of the cosmic structure formation, e.g. galaxy formation, of the matter instabilities, in obtaining the parameters of the 21 cm spectrum, and in the $\sigma_8 $ studies, respectively. These cosmological aspects  will be investigated in full detail in a future paper.}

In the present work, we have investigated the thermodynamic interpretation of the interacting dark energy-dark matter models that introduce a coupling between the components of the Universe. We have also analyzed some of the cosmological implications of this type of models.
As can be seen from the analyzes given in the text as well as in the drawing of the various diagrams, the model we examined provided acceptable predictions compared to the observations. As an important parameter in the study of cosmic evolution, the study of the Hubble parameter is of great importance. And we focused a lot on examining it, and as shown in the Figs.~\ref{fig:Hubble-rate}, \ref{fig:2d-h0}, \ref{fig:1d-likes} and \ref{fig:tri_plot} acceptable results were obtained in concordance with the observational Riess et al., and other data sets. Other important parameters that have been examined for our model are given in Tables \ref{tabpriors} and \ref{table1}.
Through our investigations we have developed some basic theoretical tools that can be used to further analyze the cosmological implications of interacting dark-energy-dark matter-ordinary matter models,  and of particle production in the early and late Universe.\\

\section*{Declaration of competing interest}
The authors declare that they have no known competing financial
interests or personal relationships that could have appeared
to influence the work reported in this paper.



\section*{Acknowledgments}
We would like to thank the anonymous referee for comments and suggestions that helped us to improve our work. H.S. would like to thank H. Firouzjahi and A. Talebian for very constructive and fruitful discussions we had on warm inflationary models.
The work of T.H. was supported by a grant of the Romanian Ministry
of Education and Research, CNCS-UEFISCDI,
project number PN-III-P4-ID-PCE-2020-2255 (PNCDI~III).

\section*{CRediT authorship contribution statement}

\textbf{T Harko: } Conceptualization, Investigation, Methodology,
Visualization, Writing - original draft, Writing - review \& editing.
\\
\textbf{K. Asadi:} Investigation, Preparing some primary parts of the original draft.
\\
\textbf{H Moshafi:} Investigation, Methodology, Observational analyzing, Visualization.
\\
\textbf{ H Sheikhahmadi:} Conceptualization, Investigation,
Methodology, Visualization, Writing - original draft, Writing- review \& editing.





\end{document}